\documentclass[fleqn,usenatbib]{mnras}

\usepackage{newtxtext,newtxmath}

\usepackage[T1]{fontenc}
\DeclareRobustCommand{\VAN}[3]{#2}
\let\VANthebibliography\thebibliography
\def\thebibliography{\DeclareRobustCommand{\VAN}[3]{##3}\VANthebibliography}

\usepackage{graphicx}	
\usepackage{amsmath}	

\usepackage{amssymb}	
\usepackage{wasysym}
\usepackage{bm}

\title[Planet formation in 55 Cancri]{On the formation of terrestrial planets between two massive planets: The case of 55 Cancri}

\author[Zhou et al.]{
Lei Zhou,$^{1,2,3}$ 
Rudolf Dvorak$^{3}$\thanks{E-mail: rudolf.dvorak@univie.ac.at (R. D.); zhouly@nju.edu.cn (L.-Y. Z.)}
and Li-Yong Zhou$^{1,2,4}$\footnotemark[1]
\\
$^{1}$School of Astronomy and Space Science, Nanjing University, 163 Xianlin Avenue, Nanjing 210046, PR China \\
$^{2}$Key Laboratory of Modern Astronomy and Astrophysics in Ministry of Education, Nanjing University, Nanjing 210046, PR China \\
$^{3}$Universit\"{a}tssternwarte Wien, T\"{u}rkenschanzstr. 17, 1180 Wien, Austria \\
$^{4}$Institute of Space Astronomy and Extraterrestrial Exploration (NJU \& CAST), PR China
}

\date{Accepted XXX. Received YYY; in original form ZZZ}

\pubyear{2015}

\begin{document}
\label{firstpage}
\pagerange{\pageref{firstpage}--\pageref{lastpage}}
\maketitle

\begin{abstract}
Considering the huge computational resources required by smoothed particle hydrodynamics (SPH) simulations and the overestimation of post-collision materials from perfect merging, we develop a statistical method to deal with collisions during the formation of planetary systems by introducing random material loss. In this method the mass and water content lost by the sole outcome from every merger vary randomly within a range dependent on the total mass and water content of colliding bodies. The application of the random loss method to the planet formation in the solar system shows a good consistency with existing SPH results. We also apply this method to the extrasolar planetary system 55 Cancri which hosts (at least) five planets and study the formation of terrestrial planets between the outermost two planets. A disk with 500 Mars mass embryos in dynamically cold orbits before the late-stage accretion phase is assumed. Scenarios with different amounts of planetary embryos and different loss parameters are adopted in our simulations. The statistical result from hundreds of simulations shows that an Earth-like planet with water inventory of roughly 6 Earth ocean could form between 55 Cnc f and d. It may reside between 1.0 and 2.6 AU but the most likely region extends from 1.5 to 2.1 AU. Thus the probability of this planet being in the potentially habitable zone (0.59--1.43 AU) is relatively low, only around 10\%. Planets 55 Cnc f and d could also be shaped and gain some water from giant impacts and consequently the orbits of them may also change accordingly.

\end{abstract}




\section{Introduction}

Many scientific investigations have been devoted during  the last thirty years to the formation of planets in our solar system and also -- since the discovery of the first planets around other stars -- to extrasolar planetary systems. Because of the diversity of their architecture many systems should be investigated separately; existing work concerning the important problem of how terrestrial planets are formed when more planets are present is very interesting \citep[see e.g.][]{2013ApJ...767...54I,2014E&PSL.392...28F,2016ApJ...821..126Q,2020A&A...634A..76B} but misses the difference in composition of the systems with respect to the number and nature of the planets and their orbits around different stars (F, G and M stars). This variety of planetary systems discovered up to now can be regarded as a good statistical sample of what is expected to be present in our Galaxy (and also other galaxies). Because we are interested in possibly habitable terrestrial planets, we need to study the presence of water in liquid form on the surface of these planets. Since planets are formed through the collision of planetesimals in early phases of a dust and gas cloud around a star, a key question in this respect is the collision behavior of celestial bodies regarding their content of water; this catastrophic event has to be modeled with specially designed effective computer codes. Several studies have shown that collision outcomes at least depend on the following parameters: the mass involved,  the mass ratio of colliding bodies, the impact velocity and the impact angle \citep[see e.g.][]{2012ApJ...745...79L,2014IAUS..310..138M,2016A&A...590A..19S,2020A&A...634A..76B}. 

This formation process happens as a long-term sequence of collisions between protoplanetary bodies of different sizes. Many evolution studies use N-body simulations to investigate the respective collisional growth to planetary embryos and to terrestrial-like planets. Most of the collision models are simplified based on perfect merging and momentum conversation \citep[see e.g.][]{2011ASL.....4..325L,2012AIPC.1468..137D,2013ApJ...767...54I,2019MNRAS.488.5604D} or they just use simple fragmentation laws \citep[see e.g.][]{1998Icar..132..113A}. \citet{2012ApJ...745...79L} analyzed the dependence of collisions on the involved mass, impact velocity, and the impact angle, which result in four different outcomes a) the efficient accretion/perfect merging, b) the partial accretion, c) the hit-and-run, and d) the disruption. The perfect merging is a very rough estimation which ignores fragmentation in partial accretion and so is the hit-and-run scenario \citep{1999Icar..142..219A,2010ChEG...70..199A}. In fact in these kinds of studies the mass of the surviving bodies are overestimated in every collision. 

\citet{2019A&A...632A..14D} proposed an N-body integrator with fragmentation and hit-and-run collisions to study the influence of embryo fragmentation on the physical and dynamical properties of terrestrial planets around solar-type stars. Their simulations suggest that the mass of habitable planets around solar-type stars is reduced by about 15\%--20\% in the presence of fragmentations.

\citet{2004ApJ...613L.157A} studied for the first time the accretion consistently by using smoothed particle hydrodynamics (SPH) computations. Later \citet{2010ApJ...714L..21K} provided SPH-based collision simulations to derive an empirical formula for the impact velocity to differentiate between merging and hit-and-run collisions. Their results led to the conclusion that only about half of the collisions lead to accretion, which agrees with the computations showing that only less than half of giant collisions result in non-merged bodies \citep{2004ApJ...613L.157A,2010ApJ...714L..21K,2012ApJ...744..137G}. 

For terrestrial planets in stable orbits inside the habitable zone (HZ) around the central star the water content is very important for possible life on the surface. Again for the simplification of the collision outcomes it is assumed in many studies that due to perfect merging the mass and water during a collision is retained \citep{2004Icar..168....1R,2013ApJ...767...54I}. But \citet{2007ApJ...666..436H} claim that in such simulations the content of water is at least 5 to 10\% overestimated (the same for the mass).  In fact \citet{2006LPI....37.2146C} found a loss of water more than 50\% for special collision velocities (1.4 times escape velocity and impact angles > 30$^\circ$).

There exists recent work \citep{2020A&A...634A..76B} where for the formation of planets in the solar system the mass and water loss during a collision was determined with SPH simulations during the numerical integration of the system. The difference in the results for the formation of terrestrial planets in the solar system between the perfect merging and SPH implemented simulations after 100 Myr is shown in detail in the respective tables and plots in \citet{2020A&A...634A..76B}. From that one can see that perfect merging is a poor approximation. The materials (especially water content) of the formed planets are significantly overestimated in perfect merging. Considering the huge computational costs of SPH simulations and the poor approximation of perfect merging, here we propose a new approach to statistically analyze the collision outcomes. 

In this paper we deal with the problem of formation of terrestrial planets between two massive planets. As a case we choose the extrasolar planetary system 55 Cancri (55 Cnc, $\rho^1$ Cancri) where (at least) five planets move around the G8 V star 55 Cnc A (see Tables~\ref{tab:55CncA} and \ref{tab:55Cnc} for more parameters). The large mass of the planets in 55 Cnc could be detected with observations via radial velocities and consequently are only minimum mass, or $M\sin{i}$ with $M$ and $i$ being the mass and inclination, respectively (except for 55 Cnc e which is also detected by transit). Nevertheless thanks to a lot of dynamics research of this system \citep[see e.g.][]{2003AsBio...3..681V,2007MNRAS.374..599R,2008ApJ...689..478R,2009Icar..201..381S,2019PASJ...71...53S}, one can estimate stability limits for the possible terrestrial planets in there.

\begin{table}
\renewcommand\arraystretch{1.2}
	\centering
	\caption{Stellar parameters of 55 Cnc A from \citet{2016A&A...586A..94L}.}
	\begin{tabular}{ccccc}
	\hline\hline
	Sp. class & Radius & Mass & Effective temp. & Luminosity \\
	\hline
	G8V & 0.96 $R_\odot$ & 0.960 $M_\odot$ & 5165 K & 0.589 $L_\odot$ \\
	\hline
	\end{tabular}
	\label{tab:55CncA}
\end{table}

One feature that makes 55 Cnc special and interesting is the similarity to our solar system, which is also the reason why we choose it: the star 55 Cnc A has a similar mass to the Sun; 55 Cnc f ($a=0.7708$ AU, $M=47.8\,M_{\oplus}$) has a similar semi-major axis $a$ to Venus ($a=0.7233$ AU, $M=0.815\,M_{\oplus}$) and 55 Cnc d ($a=5.957$ AU, $M=991.6\,M_{\oplus}$) has a similar $a$ to Jupiter ($a=5.20$ AU, $M=317.8\,M_{\oplus}$); there is a big gap between 55 Cnc f and 55 Cnc d where smaller planets may reside while we have terrestrial planets Earth, Mars and the asteroid belt between Venus and Jupiter in our solar system. The biggest difference is that the planets in 55 Cnc are much more massive than those in the solar system. But this difference is also helpful for us to understand the role of previously formed planets in the formation of terrestrial planets. No terrestrial planet has been observed between 55 Cnc f and d up to now although this big gap would allow such planets to move in stable orbits. \citet{2008ApJ...689..478R} found a large stable zone between 0.9 and 3.8 AU and demonstrated that there could be two or three additional planets in that region. In addition, \citet{2012PASJ...64...73C} suggests a possible planet around 1.5 AU in 55 Cnc and \citet{2019PASJ...71...53S} supports that an Earth-like planet could exist between 1 and 2 AU based on orbital stability simulations.

 But even if additional planets exist in 55 Cnc, how could they have formed? A lot of materials should be there in form of planetesimals or embryos during the formation or after the formation of the large planets. Very probably these large planets are Neptune to Jupiter like gas planets which formed in a very early phase like the gas giants in the solar system. What one could imagine is a scenario, like the one of the Grand Tack during the early formation of the planets in the solar system, that leads to a scattering of the planetesimals into the gap between 55 Cnc f and d due to a process of migration. The possible mean motion resonances (MMRs) between planets during the migration could further help scatter the planetesimals \citep[see e.g.][]{2004Icar..167..347K,2008IAUS..249..485Z,2020A&A...633A.153Z}. Accepting this ad hoc assumption (idea, model) we made thousands of simulations for a different number of small bodies (500 and 1500), different initial distribution of the bodies (Mars size and smaller) to model the formation of terrestrial planets between 55 Cnc f and d. This will not only answer the question about the existence of terrestrial planets in 55 Cnc, but also provide some clues about the planet formation in our solar system.

We summarize all abbreviations used in this paper in Table~\ref{tab:abbr}.
The remainder of our paper is organised as follows. We describe the model and numerical setup in Section~\ref{sec:modmet}. The comparison between SPH and our approach dealing with collisions is discussed in Section~\ref{sec:compare}. Then we apply our method to statistically study the probability of forming terrestrial planets in 55 Cnc under different scenarios and present the result in Section~\ref{sec:cnc}. Finally a summary of conclusions and a discussion come in Section~\ref{sec:condis}.

\begin{table}
	\renewcommand\arraystretch{1.2}
	\centering
	\caption{Summary of the abbreviations in this paper.}
	\begin{tabular}{cc}
	\hline\hline
	Abbreviation & Full form \\
	\hline
	HZ & habitable zone \\
	KDE & kernel density estimation \\
	MMR & mean motion resonance \\
	MMSN & minimum-mass solar nebula \\
	MVS & mixed variable symlectic \\
	PM  & perfect merging \\
	RLM & random loss method \\
	SPH & smoothed particle hydrodynamics \\
	WMF & water mass fraction \\
	\hline
	\end{tabular}
	\label{tab:abbr}
\end{table}

\section{Model and method}\label{sec:modmet}

\subsection{Initial conditions}
To investigate the formation of terrestrial planets by gravity-dominated collisions, the planetary system 55 Cnc, which hosts (at least) five planets, is assumed to be at the beginning of the late-stage accretion phase (giant impact) where the gas in the disk has dissipated and a large number of planetary embryos have formed from the materials that were scattered from inner orbits to the gap between 55 Cnc f and d due to the inward migration of giant planets. We aim to set up a general model where planets could form in the gap between two massive planets. In the case of 55 Cnc, we include planets 55 Cnc f and d in our model, which are assumed to be in their current orbits. Considering the extremely small time step required for 55 Cnc e, b and c, it is a common method for simplification to exclude the three inner planets and approximate the gravitational potential by adding their mass onto the $0.96\,M_{\astrosun}$ star. The orbital parameters of all planets obtained from \citet{2018A&A...619A...1B} are listed in Table~\ref{tab:55Cnc}. We note that although 55 Cnc is a binary star system containing a secondary star 55 Cancri B (red dwarf), the large separation ($\sim1065$ AU) of these two stars makes this companion ignorable for the dynamical evolution of orbits around 55 Cancri A \citep{1991A&A...248..485D,2003ASPC..294...43E}.

\begin{table}
\renewcommand\arraystretch{1.2}
	\centering
	\caption{Orbital elements of 5 planets in 55 Cancri from \citet{2018A&A...619A...1B}. $M$ is mass; $a$ is semi-major axis; $e$ is eccentricity; $i$ is inclination which is the angle between the line of sight and the normal to the orbital plane and $\omega$ is the argument of perihelion. Except for 55 Cnc e, the mass of other planets we use are actually the minimum mass ($M\sin{i}$) obtained from radial velocity measurements. For all planets, the longitude of ascending node and mean anomaly are set to 0.}
	\begin{tabular}{cccccr}
	\hline\hline
	Planet & $M$ $(M_{\oplus})$ & $a\ (\rm AU)$ & $e$ & $i\ (^\circ)$ & $\omega\ (^\circ)$ \\
	\hline
	55 Cnc e & 7.99  & 0.0154 & 0.05 & 6.41 & 86 \\
	55 Cnc b & 255.4 & 0.1134 & 0.00 & 0.0 & $-21.5$ \\
	55 Cnc c & 51.2  & 0.2373 & 0.03 & 0.0 & 2.4 \\
	55 Cnc f & 47.8  & 0.7708 & 0.08 & 0.0 & $-97.6$ \\
	55 Cnc d & 991.6 & 5.957  & 0.13 & 0.0 & $-69.1$ \\
	\hline
	\end{tabular}
	\label{tab:55Cnc}
\end{table}

We consider an initial debris disk with 500 Mars mass ($M_{\mars}$), or equivalently 54 Earth mass ($M_{\oplus}$), between the orbits of 55 Cnc f and d. Since the total mass of planets that have been detected in 55 Cnc ($\sim1354\,M_{\oplus}$) is much larger than that for the solar system ($\sim447\,M_{\oplus}$), it is reasonable to suppose that there is much more material in the protoplanetary disc of 55 Cnc and consequently more mass in embryos compared to the case of the solar system, where a total disk mass in solids is assumed to be $\sim6\,M_{\oplus}$ \citep{2020A&A...634A..76B} adopting the minimum-mass solar nebula (MMSN) proposed by \citet{1977Ap&SS..51..153W,1981PThPS..70...35H}.  

The theory of oligarchic growth where the embryos could accrete the mass in its feeding zone is widely adopted in the study of planet formation \citep[see e.g.][]{2000M&PS...35.1309M,2002ApJ...581..666K,2004Icar..168....1R,2008ApJ...673..502K,2020A&A...634A..76B}. The feeding zone is an annulus with width proportional to the Hill radius of an embryo or mutual Hill radius of two adjacent embryos. The so-called isolation mass which is the total mass in the feeding zone could been calculated accordingly for a specific surface density (e.g. MMSN). Therefore, given the boundary of the disk, the distribution of the embryos (mass and semi-major axis) can be uniquely determined by the isolation mass, mutual spacing and the surface density profile. 

\begin{figure}
	\centering
    \includegraphics[width=\hsize]{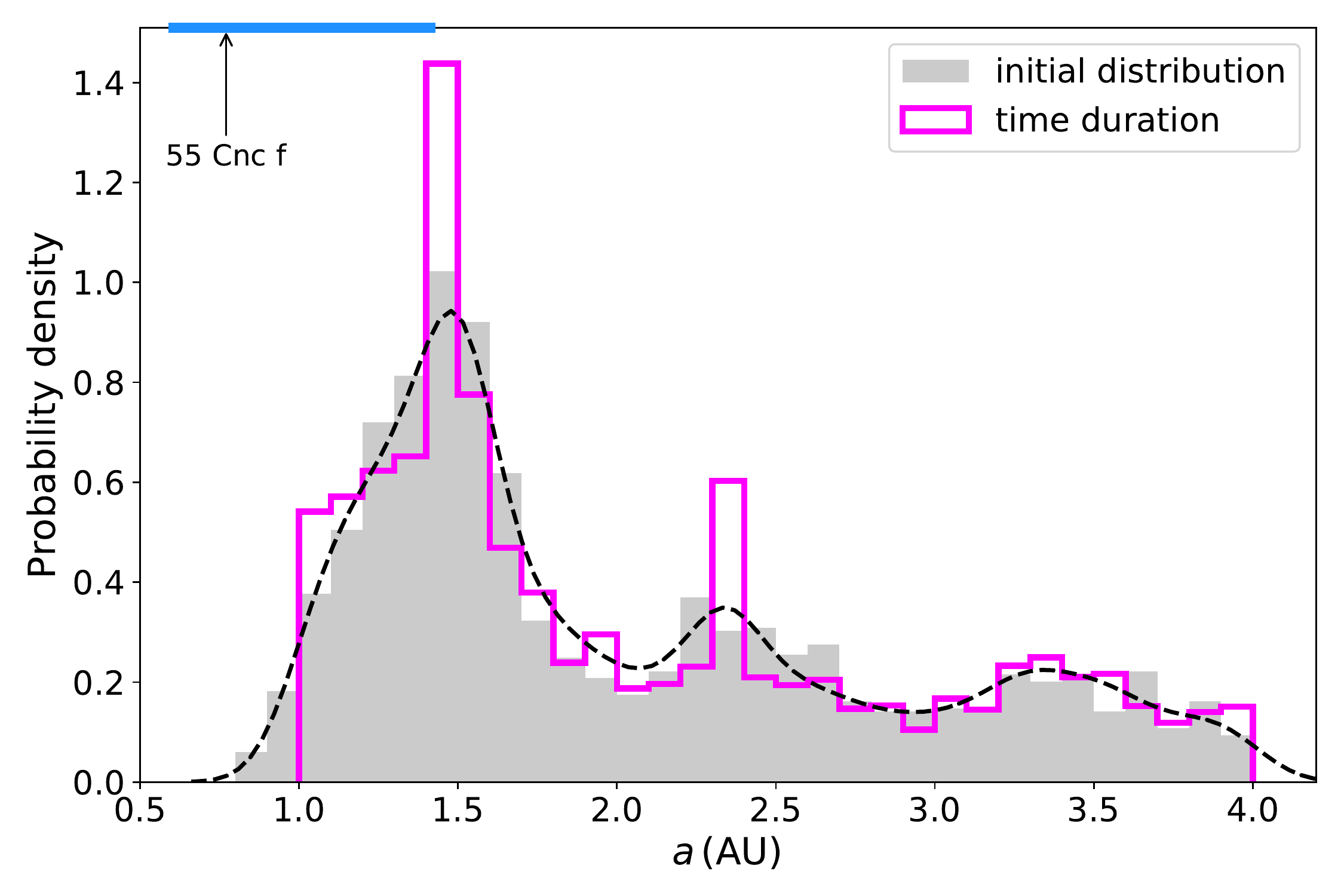}
    		\caption{Radial distribution (in probability density which means the area under the histogram is 1) of time duration (fuchsia) and initial embryos (grey). The black curve indicates the Gaussian KDE for time duration distribution. The initial distribution of embryos is obtained by randomly resampling 1500 lines of data from the KDE. The region of the potentially habitable zone (0.59--1.43 AU) is indicated by a blue bar on the top axis and the location of 55 Cnc f ($a=0.7708$ AU) is also marked.}
    		\label{fig:initdist}
	\end{figure}

However, for 55 Cnc we consider a later stage after the oligarchic growth where some massive planets have formed and migrated. We adopt another method to generate the initial distribution of embryos in this case. Under the assumption that the embryos in the gap originate from the objects that were scattered from inner orbits during the inward migration of giant planets, we randomly put one thousand massless bodies inside the orbit of 55 Cnc f and integrate the whole system including the star and all 5 planets (see Table~\ref{tab:55Cnc} for orbital parameters) for 250 kyr. Then the time duration\footnote{In particular, we record the number of time steps of massless bodies staying in each bin of semi-major axis (see Fig.~\ref{fig:initdist}) when they are scattered out to the gap between 55 Cnc f and d.} distribution of these massless bodies staying between the orbits of 55 Cnc f and d, which is also the probability distribution of a body appearing in some specific location can be calculated to mimic the radial distribution of embryos via the Gaussian kernel density estimation (KDE). We can randomly resample a dataset of any size from the calculated KDE of time duration. Therefore different initial distributions of embryos following the same KDE can be generated and one of them is shown as an example in Fig.~\ref{fig:initdist}. The objects close to the orbits of 55 Cnc f and d are very likely to be scattered away quickly. Hence, to increase the possibility of collisions and thus increase the efficiency of forming planets, we cut off the time duration distribution at 1 and 4 AU. However, there are still a few embryos extending out of the area (1--4 AU) due to the continuity of KDE. As we can see from Fig.~\ref{fig:initdist}, the embryo distribution has three peaks around 1.5, 2.3, 3.4 AU with peak values decreasing with $a$. It is supposed to be related to the MMRs with 55 Cnc f and d (see Fig.~\ref{fig:dista} in Section~\ref{subsec:finpha}). A large portion of embryos gather around the outer edge of the potentially habitable zone, which occupies the region from 0.59 to 1.43 AU \citep{1993Icar..101..108K,2013ApJ...765..131K,2014ApJ...787L..29K,2019PASJ...71...53S}.

In our work, we adopt 500 and 1500 embryos for different simulations (see Section~\ref{sec:cnc}). The initial mass of embryos follows a uniformly random distribution from 80\% to 120\% of the mean value of mass (1 and 1/3 $M_{\mars}$ for 500 and 1500 embryos, respectively). The embryos are assumed to be in dynamically cold orbits initially as a result of the dynamical friction with planetesimals and gas. Similar to the the strategy of \citet{2019MNRAS.488.5604D}, the initial eccentricity $e$ is sampled randomly between 0 and 0.08 while the inclination $i$ is between $0.5^\circ$ and $1.5^\circ$. Other angle elements are set to vary randomly between $0^\circ$ and $360^\circ$. The bulk density used to calculate the radii of these embryos is assumed to be 3 g/cm$^3$. The same value is adopted for 55 Cnc f and d, indicating the radii of roughly 0.4 and 1.1 $R_{\jupiter}$, respectively.

Considering that 55 Cnc A is a main-sequence star whose mass is very close to the solar mass \citep{2016A&A...586A..94L}, we could assume that 55 Cnc A follows a similar evolution path to the Sun and causes a similar snow line to that in the solar system at the beginning of the late-stage accretion phase. Therefore we adopt the same water mass fractions (WMF) for embryos as that used for the solar system in many studies \citep[see e.g.][]{2004Icar..168....1R,2006Icar..183..265R,2007AsBio...7...66R,2007ApJ...666..436H,2014ApJ...786...33Q,2020A&A...634A..76B}. The widely-adopted WMF varies with the semi-major axis in a piecewise function
\begin{equation}\label{eqn:wmf}
\mathrm{WMF} = \left\{ \begin{array}{lc} 10^{-5} & \qquad a < 2\,\mathrm{AU}\\ 10^{-3} & \qquad 2\,\mathrm{AU} < a < 2.5\,\mathrm{AU}\\ 0.05 & \qquad a > 2.5\,\mathrm{AU} \end{array} \right. .
\end{equation}
If we count the sum of the current inventory of surface and crustal water of the Earth, which is $\sim2.8 \times 10^{-4}\,M_{\oplus}$ \citep{1998ChGeo.145..249L}, as 1 $O_{\oplus}$, the total water mass in embryos is about 2500 $O_{\oplus}$. This value changes for different initial conditions and the difference could be up to hundreds of $O_{\oplus}$ for our 500 or 1500 embryo configuration. Each embryo outside 2.5 AU contains $\sim19\,O_{\oplus}$, which is 5000 times more than embryos within 2 AU.
	
\subsection{Collision model}

In our numerical simulations the collisions happen when two objects (including planets, protoplanets\footnote{In this paper ``protoplanet'' indicates the bodies that have accreted at least one body but not yet grown to planets.} and embryos) come closer than the sum of their radii. For perfect merging (PM hereafter) the collision is complete accretion where the outcome is one new object composed of all materials from two colliding objects. The position and velocity of the barycenter are determined as the position and velocity of the new object. However, the realistic situations are much more complicated. The SPH simulations suggest that collisions could cause different material loss according to various parameters such as impact velocity, impact angle, mass ratio and combined mass \citep[see e.g.][]{2016A&A...590A..19S,2020A&A...634A..76B}.

Although many sophisticated semi-analytical scaling laws have been developed to model collisions \citep[see e.g.][]{2009ApJ...700L.118M,2010ApJ...719L..45M,2010ApJ...714L..21K,2012ApJ...744..137G,2012ApJ...745...79L}, we propose a new method of considering a random loss of mass and water content to the sole outcome in mergers in our simulations. Chaos is the essential feature of the dynamical process and we think that the dynamical process by randomly selection of mass loss and water loss is a closer similarity of the simulation to the reality. The mass and water loss in every merger are determined by
\begin{eqnarray}
	m_L &=& m_{tot}\cdot{r_m} = m_{tot}\cdot\left[\delta_m^{low}+R_m\cdot(\delta_m^{up}-\delta_m^{low})\right],\label{eqn:mloss} \\
	w_L &=& w_{tot}\cdot{r_w} = w_{tot}\cdot\left[\delta_w^{low}+R_w\cdot(\delta_w^{up}-\delta_w^{low})\right],\label{eqn:wloss}
\end{eqnarray}
where $m_{tot}$ ($w_{tot}$) and $m_L$ ($w_L$) indicate the total and lost mass (water) in every merger. The mass (water) loss ratio $r_m$ ($r_w$) varies from $\delta_m^{low}$ ($\delta_w^{low}$) to $\delta_m^{up}$ ($\delta_w^{up}$) and is controlled by a random number $R_m$ ($R_w$) that is evaluated between 0 and 1. In practice, the water loss is included in the mass loss, so we have to make sure $w_L\leq m_L$ for each collision. 

In our statistical method, the loss ratios (especially the upper limit) should be determined by the overall comparison between results from the random loss method (RLM hereafter) and the more realistic SPH, rather than the possible loss of materials in a single collision, which could reach tens of percent \citep[see e.g.][]{2010ApJ...719L..45M,2014IAUS..310..138M,2018CeMDA.130....2B,2020IAUS..345..287B}. After a detailed comparison with SPH simulations (see Section~\ref{sec:compare}), we use $\delta_m^{low}=1\%$ and $\delta_m^{up}=8\%$ for our primary simulations. Some other values are also used to make comparisons. The parameter $\delta_w^{low}$ is always set to be the same as $\delta_m^{low}$ ($1\%$) while $\delta_w^{up}$ is set to be 2\% larger than $\delta_m^{up}$, i.e. $\delta_m^{up}+2\%$. This is because the water layer is more likely to be on or close to the surface and thus lose more in mergers. To be more realistic, we adopt the PM instead of the RLM when the mass ratio of colliding bodies exceeds 200, which could happen between planets and embryos or protoplanets which have not yet grown to at least 0.24 $M_{\oplus}$ (1/200 of the mass of 55 Cnc f). The choice of the threshold of mass ratio (200) is empirical and somewhat arbitrary. We assume the momentum to be conserved during the collisions to calculate the velocity of the post-collision object.

\subsection{Numerical setup}

In our simulations, all objects gravitationally interact with one another. We adopt various scenarios with different number of embryos (500 or 1500) or different loss ratios (see Section~\ref{subsec:lossratio}). Then the whole system is integrated for 100 Myr to make sure that the final planets are in (approximately) stable orbits. For statistical analysis, hundreds of runs of simulations are conducted for each scenario. We note that the initial conditions for different runs are different since they are obtained via random sampling.

We modify SyMBA \citep{1998AJ....116.2067D,2000AJ....120.2117L} to include the random loss in mergers. SyMBA is a second-order symplectic algorithm integrator which can handle  close encounters and collisions by switching between a variant of standard mixed variable symplectic (MVS) methods and an improved multiple time step method according to the mutual distance of objects. When there is no close encounter, SyMBA is as fast as the MVS integrator. If any collision occurs, SyMBA could decompose the mutual gravitation of the interactive pair and integrate the orbit in an iteratively reduced time step as one object approaches another. We then apply the RLM or PM to calculate the outcome of mergers once two objects touch each other. 

\section{Comparison between the random loss method and SPH}\label{sec:compare}

Before our study on the formation of terrestrial planets in 55 Cnc, we have to check the feasibility of the RLM and choose appropriate parameters for it by comparing with the SPH simulations. The comparison is based on the results published by \citet{2020A&A...634A..76B} in which the terrestrial planet formation and water delivery of the solar system were investigated via the SPH method. So we apply the RLM to the solar system with initial conditions similar to those in \citet{2020A&A...634A..76B}. The dynamical model is composed of the Sun, Jupiter and Saturn at their current location, Mars-sized embryos and Moon-sized planetesimals.

The initial semi-major axes and masses of embryos are generated simultaneously following the concept of isolation mass $M=2\pi{a}{b}\Sigma(a){R_{H,m}}$, where $\Sigma(a)$ is the surface density in MMSN and the positive parameter $b$ indicates that the separation of adjacent embryos is $b$ times the mutual Hill radii of them $R_{H,m}$. The embryo distribution can be determined uniquely for a specific disk. Here the disk is assumed to extend from 0.5 to 4 AU with a total mass of $\sim 6M_{\oplus}$, of which 70\% is allocated to embryos and 30\% is to planetesimals. The solution consists of 35 embryos with increasing mass and separations according to the heliocentric distance and 150 equal-mass planetesimals following the same surface density profile. For our simulations, we use exactly the same initial $a$ and $M$ as in \citet{2020A&A...634A..76B}. However, the eccentricity and inclination are sampled from Rayleigh distribution with scale factors $\sigma=0.005$ for embryos and $\sigma=0.05$ for planetesimals, and other angle elements are randomly distributed between 0 and $2\pi$. On the other hand, we consider the gravitational interactions and collisions between planetesimals all the time while \citet{2020A&A...634A..76B} only includes it in close encounters in their ``pp1'' model. The same WMF as equation~(\ref{eqn:wmf}) is adopted and the total water mass is about $0.1\ M_{\oplus}$, corresponding to $345\ O_{\oplus}$. Besides, we use the same initial conditions for planets Jupiter and Saturn in eccentric orbits. 

 We refer to the scenarios ``SPH11-eJS-pp1'' to ``SPH15-eJS-pp1'' in \citet{2020A&A...634A..76B} for comparisons and rename them ``SPH1'' to ``SPH5'' in this paper. We also generate 5 different initial conditions and denote them by ``RLM1'' to ``RLM5''. To obtain statistical results, we carry out 20 runs for each initial conditions with the RLM incorporated in SyMBA. The mass loss parameters $\delta_m^{low}$ and $\delta_m^{up}$ are chosen to be 1\% and 8\% while the water loss parameters $\delta_w^{low}$ and $\delta_w^{up}$ are 1\% and 10\% after plenty of test simulations. We note that a slightly different $\delta_w^{up}$ does not have a remarkable influence on the final water content in formed planets. For example, the differences in the total water mass of formed planets are only $+2.9\%$ and $-2.8\%$ for $\delta_w^{up}=8\%$ and $\delta_w^{up}=12\%$, respectively. To better verify the improvement of the RLM, we also refer to the results adopting PM from \citet{2020A&A...634A..76B}. The corresponding scenario names ``PM07-eJS-pp1'' to ``PM09-eJS-pp1'' are replaced by ``PM1'' to ``PM3'' in this paper. The whole system is integrated for hundreds of millions of years to make sure all remaining bodies are in (approximately) stable orbits.

\begin{table*}
\renewcommand\arraystretch{1.2}
	\centering
	\caption{Results of simulations obtained by SPH, PM \citep{2020A&A...634A..76B} and RLM for the solar system. Symbol `$N$' indicates the (average) number for a specific quantity while `$M$' and `$W$' represent (average) mass and water mass, respectively. The subscript `pl' means formed planet; `meg' indicates merging (among embryos and protoplanets); `dis' represents the discarded materials including those being ejected from the system or colliding with the planets (Jupiter and Saturn) or the Sun. `p-h' (given in brackets) refers to the respective quantity only within the potentially habitable zone (0.75--1.5 AU). Mass is given in $M_{\oplus}$ and water is given in $O_{\oplus}$. Mean values and corresponding standard deviations for all scenarios are highlighted in bold. The outlier of $W_{\rm pl}$ in SPH4 is marked with a box. Since \citet{2020A&A...634A..76B} does not provide the mass of water that was ejected from the system or collided with the giant planets for each scenario (but mean values), we do not have $W_{\rm dis}$ in this table for SPH and PM. And the standard deviation for $M_{\rm dis}$ which is marked with asterisk only counts the mass of water that collided with the Sun.}
	\begin{tabular}{cllllll}
	\hline\hline
	Scenario & $N_{\rm pl}$ (p-h) & $M_{\rm pl}$ (p-h) & $W_{\rm pl}$ (p-h) & $M_{\rm meg}\ (W_{\rm meg})$ & $M_{\rm dis}\ (W_{\rm dis})$ \\
	\hline
	SPH1 & 3 (2) & 2.00 (1.86) & 0.52 (0.52) & 1.04 (2.35) & 3.05 \\
	SPH2 & 2 (1) & 1.10 (1.00) & 25.8 (25.8) & 1.82 (7.3) & 3.16 \\
	SPH3 & 2 (1) & 1.07 (0.83) & 19.7 (19.4) & 1.46 (51) & 3.56 \\
	SPH4 & 2 (0) & 1.71 (0.0)  & \boxed{98\ (0.0)}    & 1.20 (9.2) & 3.18 \\
	SPH5 & 3 (1) & 1.48 (1.03) & 2.38 (1.59) & 1.31 (8.1) & 3.30 \\
		 & \boldsymbol{$2.40\pm0.49$} & \boldsymbol{$1.47\pm0.36$} & \boldsymbol{$29.28\pm35.71$}   & \boldsymbol{$1.4\pm0.3$} & \boldsymbol{$3.25\pm0.17$} \\
		 & \boldsymbol{$(1.00\pm0.63)$} & \boldsymbol{$(0.94\pm0.59)$} & \boldsymbol{$(9.5\pm10.9)$}   & \boldsymbol{$(16\pm18)$} & \boldsymbol{$(300\pm17$}$^*)$ \\
	\hline
	RLM1 & 1.65 (1.10) & 1.44 (1.04) & 7.80 (4.56) & 0.46 (3.71) & 4.19 (334)  \\
	RLM2 & 2.10 (1.00) & 1.58 (0.84) & 17.23 (7.59) & 0.44 (4.01) & 4.06 (324) \\
	RLM3 & 1.55 (0.95) & 1.21 (0.88) & 3.23 (1.63) & 0.37 (2.49) & 4.52 (339) \\
	RLM4 & 1.65 (0.95) & 1.57 (1.00) & 6.73 (1.58) & 0.48 (4.76) & 4.05 (334) \\
	RLM5 & 2.00 (0.80) & 1.58 (0.86) & 14.61 (3.24) & 0.43 (2.54) & 4.07 (328) \\
	    & \boldsymbol{$1.79\pm0.22$} & \boldsymbol{$1.48\pm0.14$} & \boldsymbol{$9.92\pm5.19$} & \boldsymbol{$0.44\pm0.04$} & \boldsymbol{$4.18\pm0.18$} \\
	    & \boldsymbol{$(0.96\pm0.10)$} & \boldsymbol{$(0.92\pm0.08)$} & \boldsymbol{$(3.72\pm2.23)$} & \boldsymbol{$(3.50\pm0.88)$} & \boldsymbol{$(332\pm5)$} \\
	\hline
	PM1 & 3 (1) & 2.96 (1.83) & 34 (0.064) & 0 & 3.13 \\
	PM2 & 2 (1) & 2.76 (0.91) & 2.8 (2.7)  & 0 & 3.33 \\
	PM3 & 4 (2) & 3.34 (2.30) & 3.4 (3.32) & 0 & 2.75 \\
		 & \boldsymbol{$3.00\pm0.82$} & \boldsymbol{$3.02\pm0.24$} & \boldsymbol{$13.40\pm14.57$}   & 0 & \boldsymbol{$3.07\pm0.24$} \\
		 & \boldsymbol{$(1.33\pm0.47)$} & \boldsymbol{$(1.68\pm0.58)$} & \boldsymbol{$(2.03\pm1.41)$}   & 0 & \boldsymbol{$(332\pm30$}$^*)$ \\
	\hline
	\end{tabular}
	\label{tab:comp}
\end{table*}

The results from these three methods are summarized in Table~\ref{tab:comp}. As \citet{2020A&A...634A..76B} did, we only count remaining protoplanets at the end of integration as formed planets. In SPH simulations, the collisions could produce either 0, 1 or 2 post-collisional bodies (total disruption, accretion or erosion, hit-and-run) while the RLM and PM always generate only one for each collision. Always merging to one body could make it easier to form large planets. It is reflected by the average mass of formed planets ($M_{\rm pl}/N_{\rm pl}$), which is 0.61 $M_{\oplus}$ for SPH, 0.83 $M_{\oplus}$ for the RLM and 1.00 $M_{\oplus}$ for PM. \citet{2020A&A...634A..76B} have demonstrated that among all collisions (150--200) for each scenario in their simulations, more than half (about 100) result in hit-and-run, in which the number of bodies stay unchanged. Hence the frequent occurrence of hit-and-run could relieve the decline of the number of bodies, leading to more number of bodies, and thus more collisions. In our simulations, there are about 40 collisions for each run, which is only about 20\%--27\% of that in SPH. And as we can see obviously from Table~\ref{tab:comp}, despite similar mass of (potentially habitable) formed planets, we have much less material loss in mergers which is removed from the system after collisions (0.44 $M_{\oplus}$ v.s. 1.4 $M_{\oplus}$ in $M_{\rm meg}$ and 3.50 $O_{\oplus}$ v.s. 16 $O_{\oplus}$ in $W_{\rm meg}$). However, if we normalize the loss in mass and water by the number of collisions (we take 40 for the RLM and 150--200 for SPH), we can get similar average values of material loss per collision for the RLM and SPH. For SPH, these values are 0.0070--0.0093 $M_{\oplus}$ for mass loss and 0.080--0.11 $O_{\oplus}$ for water loss while for the RLM, they are 0.011 $M_{\oplus}$ and 0.088 $O_{\oplus}$, respectively. Taking account of the large uncertainty of $M_{\rm meg}$ (0.3 $M_{\oplus}$), the mass loss per collision in our simulations is still within the range of that for SPH (0.0055--0.0113 $M_{\oplus}$).  

The total mass of all formed planets and of planets only in the potentially habitable zone are consistent perfectly with each other for the RLM and SPH (1.48 $M_{\oplus}$ v.s. 1.47 $M_{\oplus}$, 0.92 $M_{\oplus}$ v.s. 0.94 $M_{\oplus}$). On the average, only 1.79 planets form in each run with the RLM, of which 0.96 is in the potentially habitable zone. For SPH, more planets form (2.40) but the number of planets in the potentially habitable zone (1.00) is almost the same as that in our simulations. Although for the RLM each planet could gain more mass on average, but the mean mass of potentially habitable planets, which are 0.96 and 0.94 $M_{\oplus}$ for the RLM and SPH, respectively, are very similar and close to the Earth mass. 

\begin{figure}
	\centering
    \resizebox{\hsize}{!}{\includegraphics{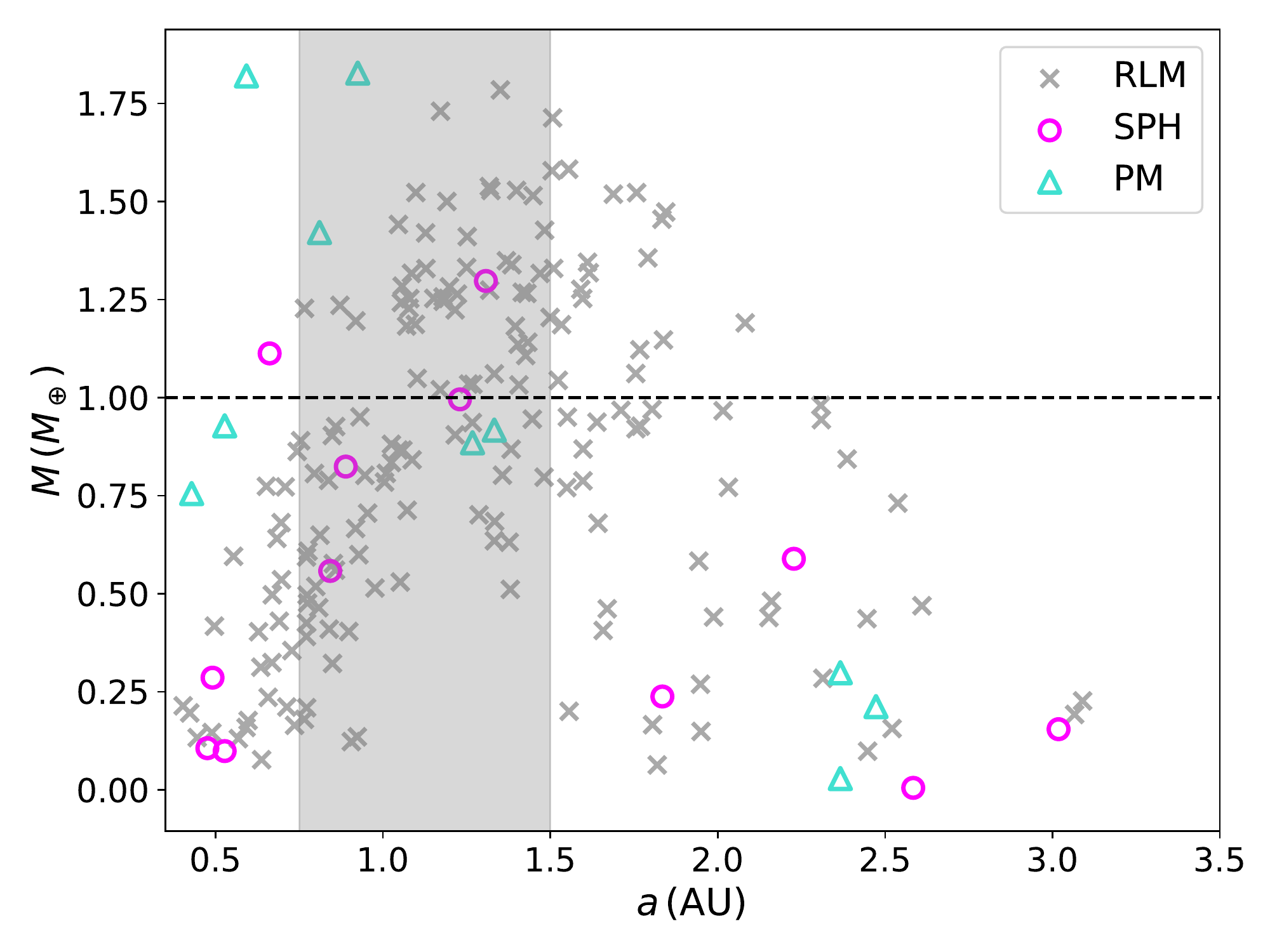}}
    		\caption{Mass distribution as a function of semi-major axis for the RLM, SPH and PM. The data for SPH and PM are taken from Fig. 4 of \citet{2020A&A...634A..76B}. The region of the potentially habitable zone for the solar system (0.75--1.5 AU) is depicted by a grey shadow. The horizontal line indicates $M=1\,M_{\oplus}$.}
    		\label{fig:mdistcomp}
	\end{figure}

Combining the outcome from all simulation runs, we present the mass distributions of formed planets as a function of semi-major axis in Fig.~\ref{fig:mdistcomp}. The results from our simulation with the RLM agree well with that from SPH \citep{2020A&A...634A..76B}. Due to the perturbations from Jupiter and Saturn, the mass drops rapidly from 1.5 AU. The most massive planets reside around 1.35 AU for the RLM and SPH. For the RLM, the maximum mass of formed planets is $\sim 1.78\,M_{\oplus}$ while it is $~\sim1.30\,M_{\oplus}$ for SPH. The PM could produce larger planets with mass up to $\sim1.83\,M_{\oplus}$ but in orbits with semi-major axis around 0.75 AU, which is much smaller compared to the cases of the RLM and SPH.

As is shown in Table~\ref{tab:comp}, the water mass in formed planets is underestimated in our simulations by $\sim66\%$ (9.92 $O_{\oplus}$ v.s. 29.28 $O_{\oplus}$). Moreover, the RLM brings in less water to the potentially habitable zone (3.72 $O_{\oplus}$ v.s. 9.5 $O_{\oplus}$). The reason is that the radial mixing is less sufficient and less effective in our simulations considering that the water-rich bodies all reside in the outer disk ($a>2.5$ AU) initially. However, the deviations in the water content are very large between different scenarios and different simulation runs, especially for SPH. Hence the values of water mass in formed planets considering the uncertainties are more or less consistent for two methods. And if we exclude the extremely large value of $W_{\rm pl}$ in scenario SPH4 (98 $O_{\oplus}$), the mean value of $W_{\rm pl}$ for SPH becomes 12.1 $O_{\oplus}$ and is very close to our result. We discard the materials that either collided with the giant planets, the Sun or was ejected from the system during the integration. More materials are removed from the system for the RLM.

The complete accretion could always result in more planets and larger mass than the other two methods (see Table~\ref{tab:comp}). So the possibility and efficiency of forming planets are heavily overestimated for PM. Despite no water loss in collisions with PM, the water delivery based on the radial mixing is still not as effective as that in SPH, leading to lower water content in the formed planets. Considering the sole outcome of mergers in both RLM and PM, we suppose that hit-and-run could help drive more protoplanets, embryos and planetesimals inward and accordingly transport more materials to the potentially habitable zone.

Overall, even the SPH simulations do not obtain convergent results considering the large deviations between different scenarios. The RLM could generate statistical results with much smaller deviations based on numerous simulation runs and the comparison has confirmed that the results are reliable and are much better than that from PM. Moreover, it is as simple as PM and could save a lot of computation resource. However, this method is not as ``accurate'' as SPH, especially when you just have a few runs. The statistical method based on the RLM is effective and promising to deal with the collisions and formation of planets.

\section{Statistical study of 55 Cancri}\label{sec:cnc}

In this section we study the formation of terrestrial planets in the extrasolar planetary system 55 Cnc. The RLM is applied to conduct a statistical analysis of the formation probability and characteristics of possible terrestrial planets between 55 Cnc f and d. We adopt various scenarios with different loss ratios and different number of embryos. The scenarios and corresponding results are summarized in Table~\ref{tab:mod}. For two primary scenarios A (500 embryos) and B (1500 embryos), we use mass loss ratio from 1\% to 8\% and water loss ratio from 1\% to 10\%, the same as what we use for the solar system (see Section~\ref{sec:compare}). 300 simulation runs with different initial conditions are carried out for scenarios A and B. All of the other scenarios (C, D, E, F) contain 500 embryos, but with different loss ratios and 100 runs are conducted for each of them. The PM, which could be regarded as a special case of the random loss with all loss ratios equal to zero, is also included for comparisons. 

We define the remaining protoplanets whose mass exceed $0.2\,M_{\oplus}$ at the end of integration (100 Myr) as newly formed planets and the rest bodies (if any) are ignored\footnote{The neglected bodies have a total mass of $\sim0.03\,M_{\oplus}$ on average for each run, accounting for only 0.06\% of the initial disk.}. The characteristics including the location, material distribution, and orbital parameters of the formed planets, especially those residing in the potentially habitable zone (0.59--1.43 AU) are analyzed. We also inspect the growing process of the planets via collisions, which is also the process of accretion and water transport. In the end, we discuss how different loss parameters influence the formation of terrestrial planets.

\begin{table*}
\renewcommand\arraystretch{1.2}
		\centering
	\caption{Summary of parameters and results for different scenarios. $N_{\rm em}$ is the amount of initial planetary embryos; $\delta_m^{low}$, $\delta_m^{up}$, $\delta_w^{low}$ and $\delta_w^{up}$ are parameters for mass and water loss (see equations~(\ref{eqn:mloss}) and (\ref{eqn:wloss})). Symbol `$N$' indicates the average number for a specific quantity while `$M$' and `$W$' represent mass and water mass, respectively. The subscript `pl' means formed planet; `meg' indicates merging (among embryos and protoplanets) while `ejc' indicates ejection from the system; `cf' represents collision with 55 Cnc f while `cd' represents collision with 55 Cnc d. `p-h' refers to the respective quantity only within the potentially habitable zone (0.59--1.43 AU). Mass is given in $M_{\oplus}$ and water is given in $O_{\oplus}$.}
	\begin{tabular}{cccclllllllll}
	\hline\hline
	Scenario & $N_{\rm em}$ & $\delta_w^{low}$ & $\delta_w^{up}$ & $N_{\rm pl}$ & $M_{\rm pl}$ & $W_{\rm pl}$ & $N_{\rm meg}$ & $N_{\rm cf}$ & $M_{\rm meg}$ & $M_{\rm ejc}$ & $M_{\rm cf}$ & $M_{\rm cd}$ \\
	 & & $(\delta_m^{low})$ & $(\delta_m^{up})$ & (p-h) & (p-h) & (p-h) & $(N_{\rm ejc})$ & $(N_{\rm cd})$ & ($W_{\rm meg}$) & ($W_{\rm ejc}$) & ($W_{\rm cf}$) & ($W_{\rm cd}$)\\
	\hline
	A & 500  & 1\% & 8\% & 1.01 & 1.086 & 5.79 & 118 & 38 & 1.75 & 44.96 & 4.98 & 0.75 \\
	  &      & 1\% & 10\% & 0.11 & 0.0969 & 0.62 & 336 & 6 & 15.7 & 2384 & 47.9 & 72.5 \\
	\hline
	B & 1500 & 1\% & 8\% & 1.14 & 1.058 & 4.39 & 497 & 86 & 2.89 & 44.75 & 4.06 & 0.75 \\
	  &      & 1\% & 10\% & 0.19 & 0.145 & 0.65 & 898 & 17 & 20.8 & 2397 & 41.4 & 76.5 \\
	\hline
	C & 500  & 1\% & 5\% & 1.10 & 1.696 & 8.04 & 119 & 41 & 1.29 & 44.39 & 5.37 & 0.76 \\
	  &      & 1\% & 7\% & 0.17 & 0.208 & 0.37 & 333 & 6 & 12.1 & 2368 & 50.6 & 71.3 \\
	\hline
	D & 500  & 1\% & 10\% & 1.03 & 0.902 & 5.04 & 117 & 37 & 2.03 & 45.16 & 4.75 & 0.72\\
	  &      & 1\% & 12\% & 0.13 & 0.0963 & 0.31 & 338 & 6 & 18.5 & 2389 & 51.1 & 68.7 \\
	\hline
	E & 500  & 1\% & 15\% & 0.89 & 0.534 & 2.15 & 112 & 35 & 2.54 & 45.35 & 4.25 & 0.80 \\
	  &      & 1\% & 17\% & 0.20 & 0.0958 & 0.52 & 346 & 6 & 21.6 & 2446 & 45.1 & 68.6 \\
	\hline
	PM & 500 & 0 & 0 & 1.23 & 2.525 & 16.7 & 117 & 44 & 0 & 44.09 & 6.14 & 0.71 \\
	  &      & 0 & 0 & 0.38 & 0.783 & 5.10 & 331 & 6 & 0 & 2404 & 51.6 & 70.7 \\
  	\hline
	\end{tabular}
	\label{tab:mod}
\end{table*}

\subsection{Final phase of the planetary system}\label{subsec:finpha}

After an integration of 100 Myr, the planetary systems reach (approximately) stable configurations with 0/1/2/3 new planets formed via giant impacts. On average, there are 1.01 and 1.14 formed planets in each simulation run of scenarios A and B, of which 0.11 and 0.19 are in the potentially habitable zone (0.59--1.43 AU), as we can see from Table~\ref{tab:mod}. This means that although an extra Earth-like planet could exist in 55 Cnc, the probability of this planet being in the potentially habitable zone is only 11\% and 17\%, based on scenarios A and B, respectively. Due to the larger number of embryos in scenario B, more planets are expected to form. But at the same time, more embryos means more collisions (600 v.s. 162), and thus more mass loss (2.89 $M_{\oplus}$ v.s. 1.75 $M_{\oplus}$), which would reduce the materials for planet formation. Besides, the lower individual mass of embryos in scenario B also makes it more difficult to form a massive planet by accretion. Nevertheless, considering that less mass collides with planets in scenario B (4.81 $M_{\oplus}$ v.s. 5.73 $M_{\oplus}$) while the ejected mass is comparable for these two scenarios, finally the mass of formed planets is just slightly lower (1.058 $M_{\oplus}$) than that in scenario A (1.086 $M_{\oplus}$).

\begin{figure}
	\centering
    \resizebox{\hsize}{!}{\includegraphics{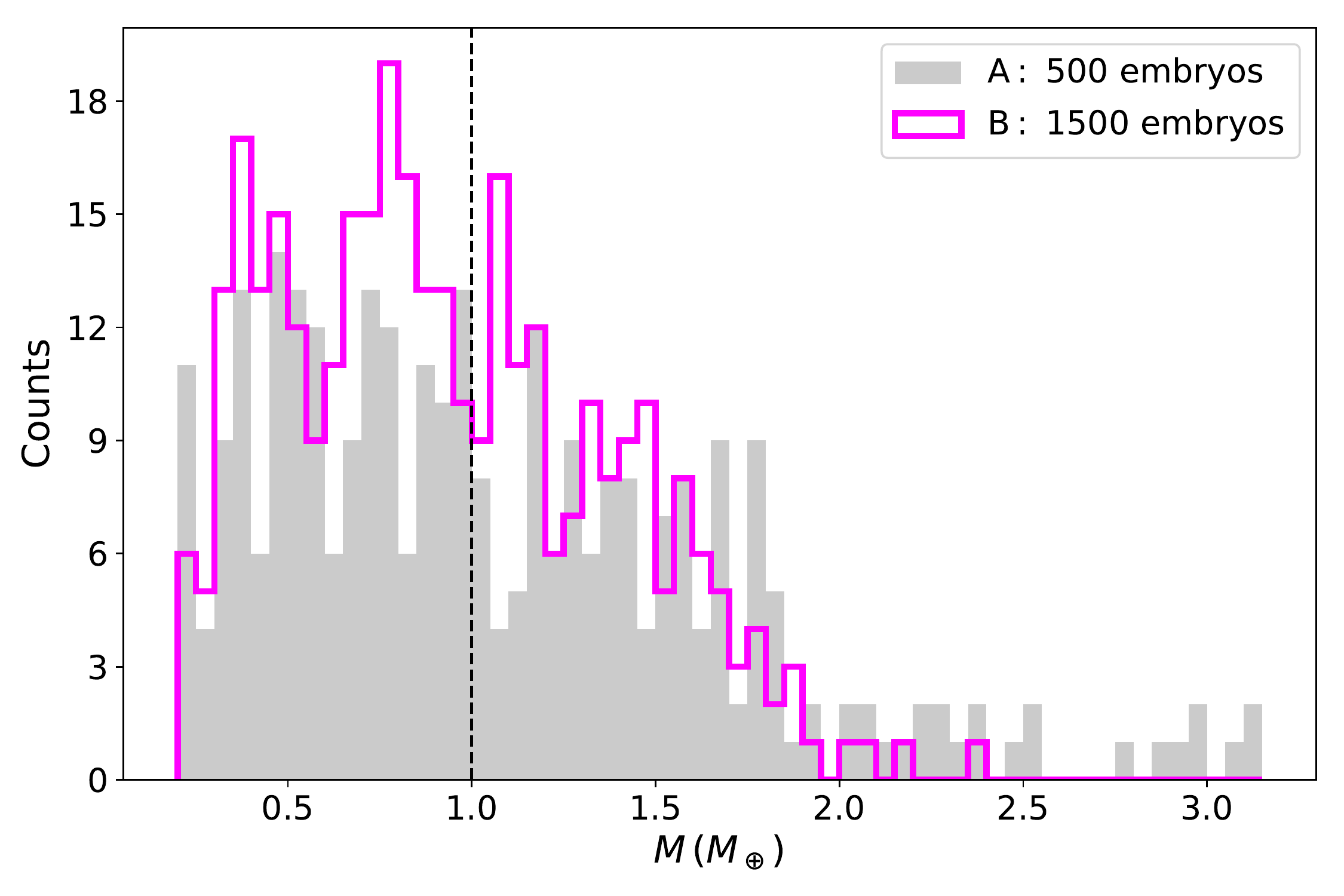}}
    		\caption{Final mass distribution of formed planets for scenarios A (500 embryos, grey) and B (1500 embryos, fuchsia) from all 300 runs. The dark line indicates $M=1\,M_{\oplus}$.}
    		\label{fig:distm}
	\end{figure}
	
Furthermore, the mass distribution of formed planets shown in Fig.~\ref{fig:distm} supports that there exist more small planets in scenario B than in scenario A. In fact, 59\% of the formed planets in scenario B are less massive than the Earth, and for scenario A this value is 53\%. A disk with fewer but larger embryos could result in more massive planets. The most massive planets are 3.25 $M_{\oplus}$ and 2.4 $M_{\oplus}$ for scenarios A and B, respectively. 

\begin{figure}
	\centering
    \resizebox{\hsize}{!}{\includegraphics{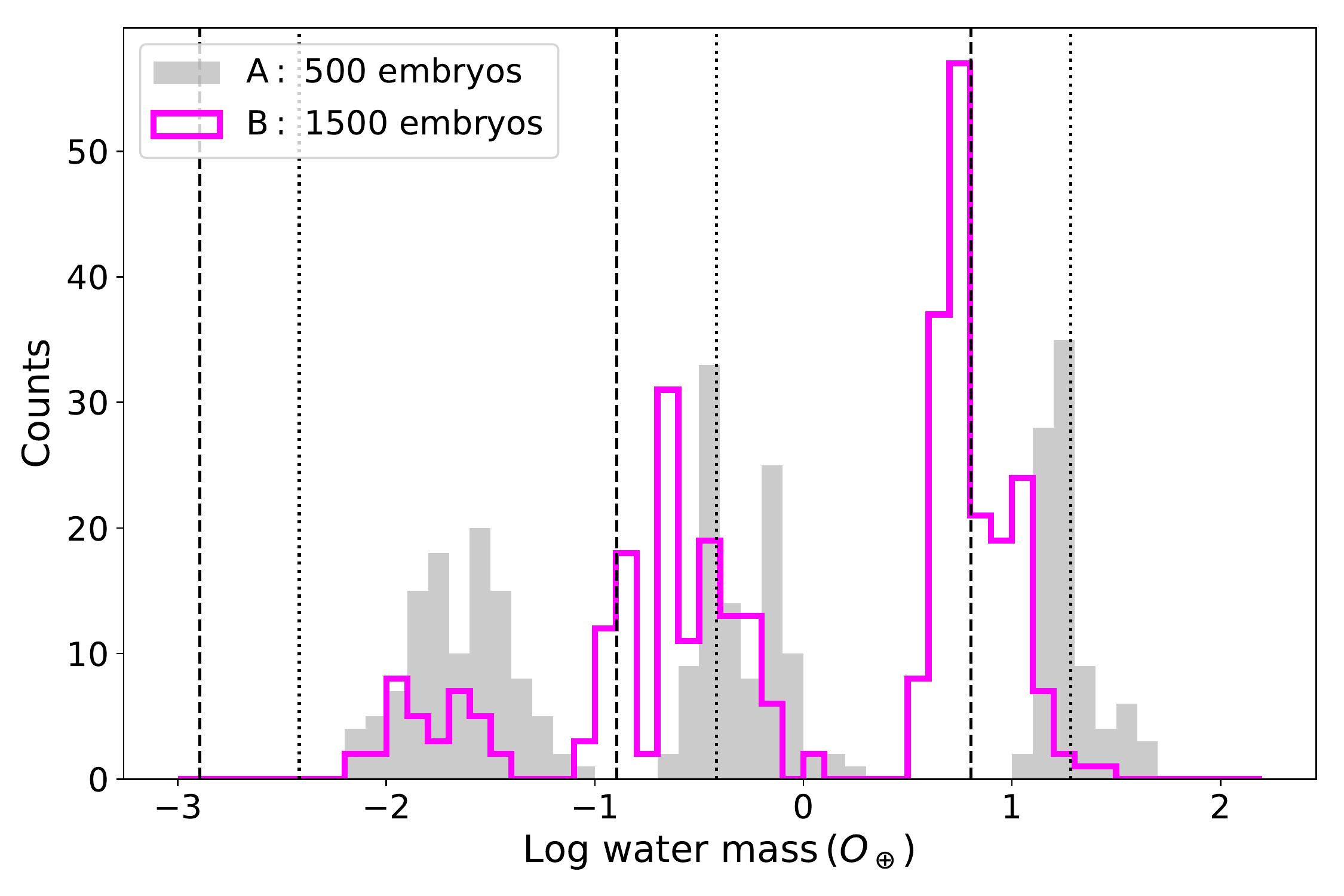}}
    		\caption{Final water mass (in $O_{\oplus}$) distribution of formed planets for scenarios A (500 embryos, grey) and B (1500 embryos, fuchsia) from all 300 runs. The dotted (dashed) lines indicate different water mass for a 1 (or 1/3) $M_{\mars}$ body with WMF according to equation~(\ref{eqn:wmf}). We note the log $x$-axis.}
    		\label{fig:distw}
	\end{figure}
	
From Table~\ref{tab:mod} we can also see that only $\sim0.2\%$ of the initial water content ($\sim2500\,O_{\oplus}$) are reserved. The ejected embryos take away more than 95\% of the water from the planetary system. Also more water content is found in scenario A (5.79 $O_{\oplus}$ v.s. 4.39 $O_{\oplus}$). The final water mass distribution is shown in Fig.~\ref{fig:distw}, from which we can see three separate regimes. The water inventory of the post-collision object is determined by the total water content of colliding bodies and also the water loss in collisions. As we know from equation~(\ref{eqn:wmf}), the initial water distribution is dependent on the location of embryos. For planets formed by only accreting water-poor embryos from the inner disk ($a<2\,{\rm AU}$), the final water content must be higher than the initial inventory of one such embryo, but still at a low level due to the limitation of the number of collisions. This corresponds to the regime around $10^{-1.7}\,(\approx0.02)$ $O_{\oplus}$ in Fig.~\ref{fig:distw}. But as long as water-rich embryos beyond 2 or 2.5 AU participate, the final water inventory will increase significantly. Even if it collides with water-poor embryos or protoplanets every time, one water-rich embryo could retain about $0.95^{15}\approx46.3\%$ of the initial water content and this causes the left edge of the third regime. Here we use a typical value of water loss 5\% and assume 15 collisions, which is already a large number of mergers to form one planet (see Section~\ref{subsec:accre}). Due to the low amount of embryos from 2.0 AU to 2.5 AU (see Fig.~\ref{fig:initdist}) and insufficient radial mixing of embryos and protoplanets, these three regimes are separate and have no overlaps, which could clearly point out the different origins of embryo components of formed planets. 

Compared to scenario A (30\%), there are more water-rich planets with at least 1 $O_{\oplus}$ in scenario B (52\%) because more water-rich embryos have the chance to be scattered inward and then accreted by embryos or protoplanets. But due to less initial inventory of individual embryo and more collision loss, the overall distribution of water mass is smaller for scenario B. In general, the final water content for a single run is very uncertain for both scenarios, depending heavily on the fact that if there are any water-rich embryos involved in the growing process of formed planets.

\begin{figure}
	\centering
    \includegraphics[width=\hsize]{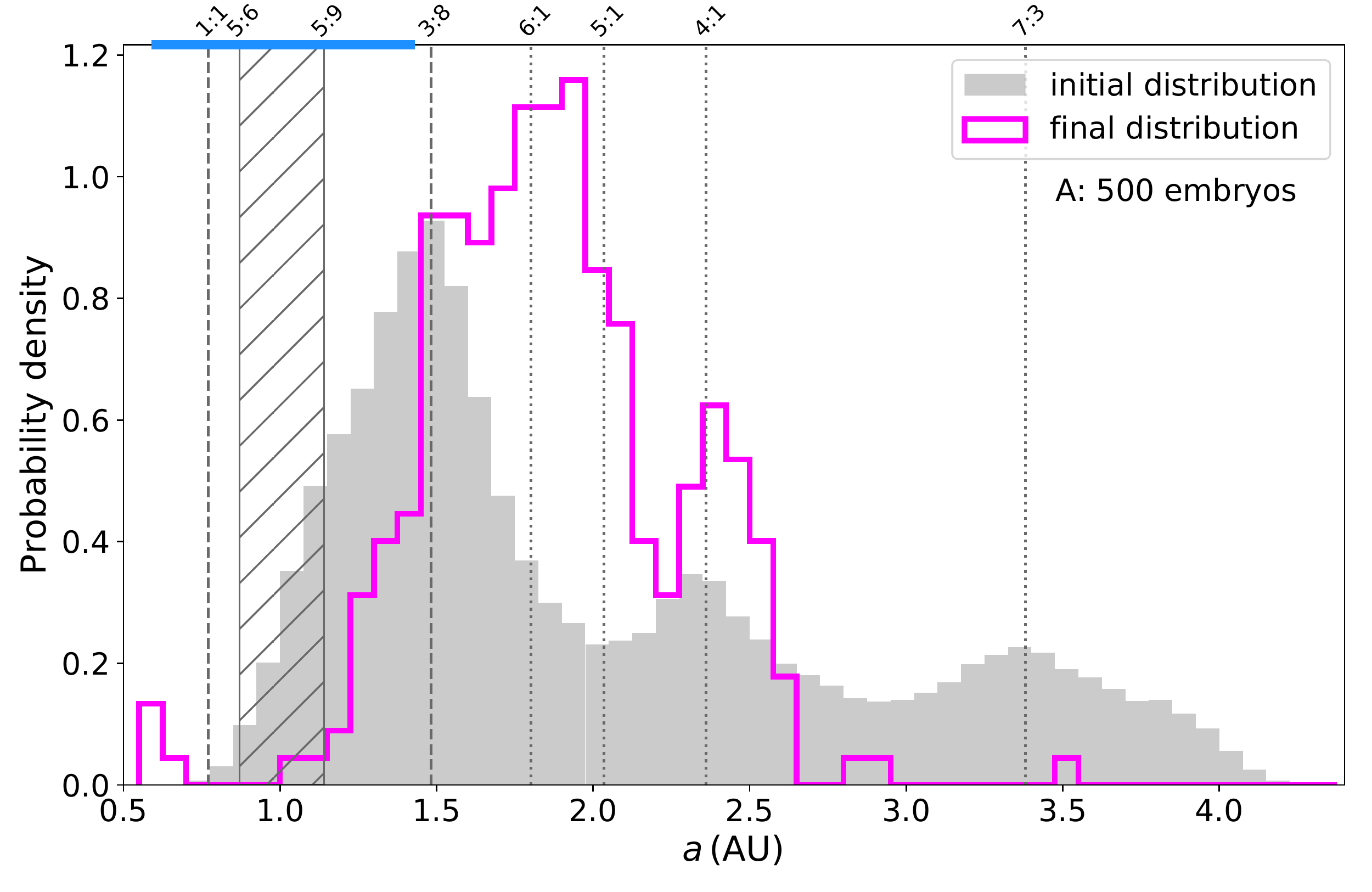}
    \includegraphics[width=\hsize]{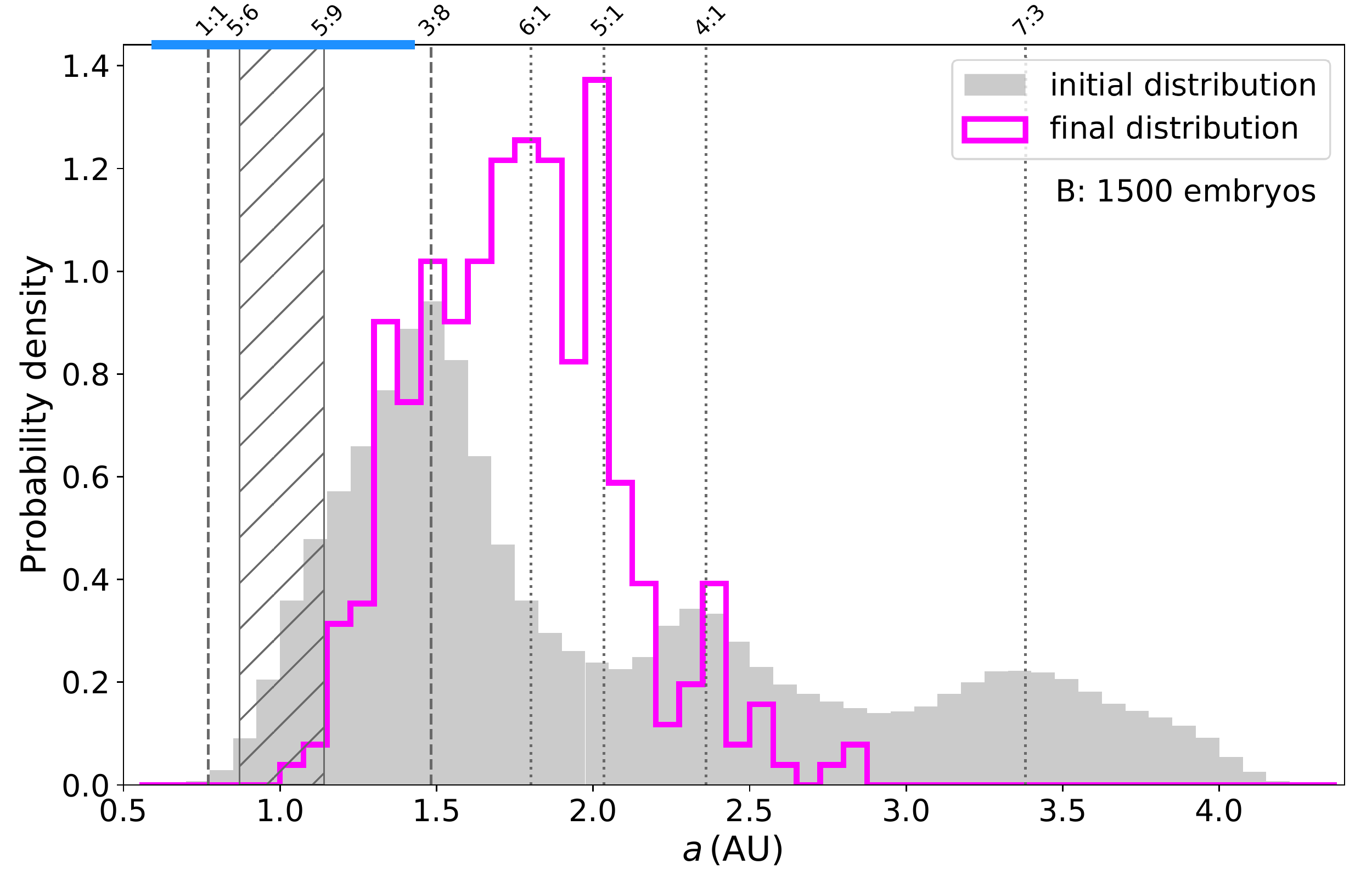}
    		\caption{Radial distribution (in probability density) of initial embryos (grey) and final planets (fuchsia). The \textit{upper} panel is for scenario A (500 embryos) while the \textit{lower} panel is for scenario B (1500 embryos). The data from all runs (300) of the same scenario are combined. The dashed lines indicate the outer MMRs with (initial) 55 Cnc f while the dotted lines indicate the inner MMRs with (initial) 55 Cnc d. The shadow represents the range covering plenty of MMRs from 5:6 to 5:9 with (initial) 55 Cnc f. The region of the potentially habitable zone (0.59--1.43 AU) is indicated by a blue bar on the top axis. We note that the location of 1:1 MMR with (initial) 55 Cnc f also indicates the semi-major axis of this planet.}
    		\label{fig:dista}
	\end{figure}

Unlike the mass and water content, the locations of formed planets are less related to the size and compositions of embryos, but more related to the dynamics of the system. The final radial distributions (see Fig.~\ref{fig:dista}) are apparently different from the initial distribution of embryos. Initially the embryos gather around the locations of the 3:8 MMR (1.5 AU) with 55 Cnc f and 4:1 (2.3 AU), 7:3 (3.4 AU) MMRs with 55 Cnc d. But the final distributions reveal that most newly formed planets are located between 1.0 and 2.6 AU for both 500 and 1500 embryos simulations. The peak which represents the most likely region of reserving formed planets is between 1.5 and 2.1 AU. There are no strong MMRs with either 55 Cnc f or d in this region. We also note another peak appears around 2.36 AU, where the 4:1 MMR with 55 Cnc d is located. It is worth noting that some planets could also move inward and cross the orbit of 55 Cnc f in scenario A. At the beginning of giant impacts, the 3:1 MMR with 55 Cnc d could quickly clear nearby orbits and hinder the inward movement of outer embryos. Moreover, there are many outer MMRs with 55 Cnc f which are very close to each other such as 5:6, 4:5, 7:9, 3:4, 5:7, 2:3, 3:5, 4:7, 5:9 MMRs (see the shadow region in Fig.~\ref{fig:dista}). The possible resonance overlapping could also destabilize the orbits and prevent the formation of planet there. On the other hand, the gravitational perturbation from the embryo disk could drive two planets approach each other. Moreover, the exchange of orbital angular momentum with the disk could also cause the radial migration of planets \citep{1984Icar...58..109F,1996P&SS...44..431F,1999AJ....117.3041H}. Like Jupiter, 55 Cnc d could also be an efficient ejector which loses orbital angular momentum during the interaction with embryos and thus moves inward. As a result, the locations of MMRs with these planets change accordingly and sweep more regions.

\begin{figure}
	\centering
     \includegraphics[width=\hsize]{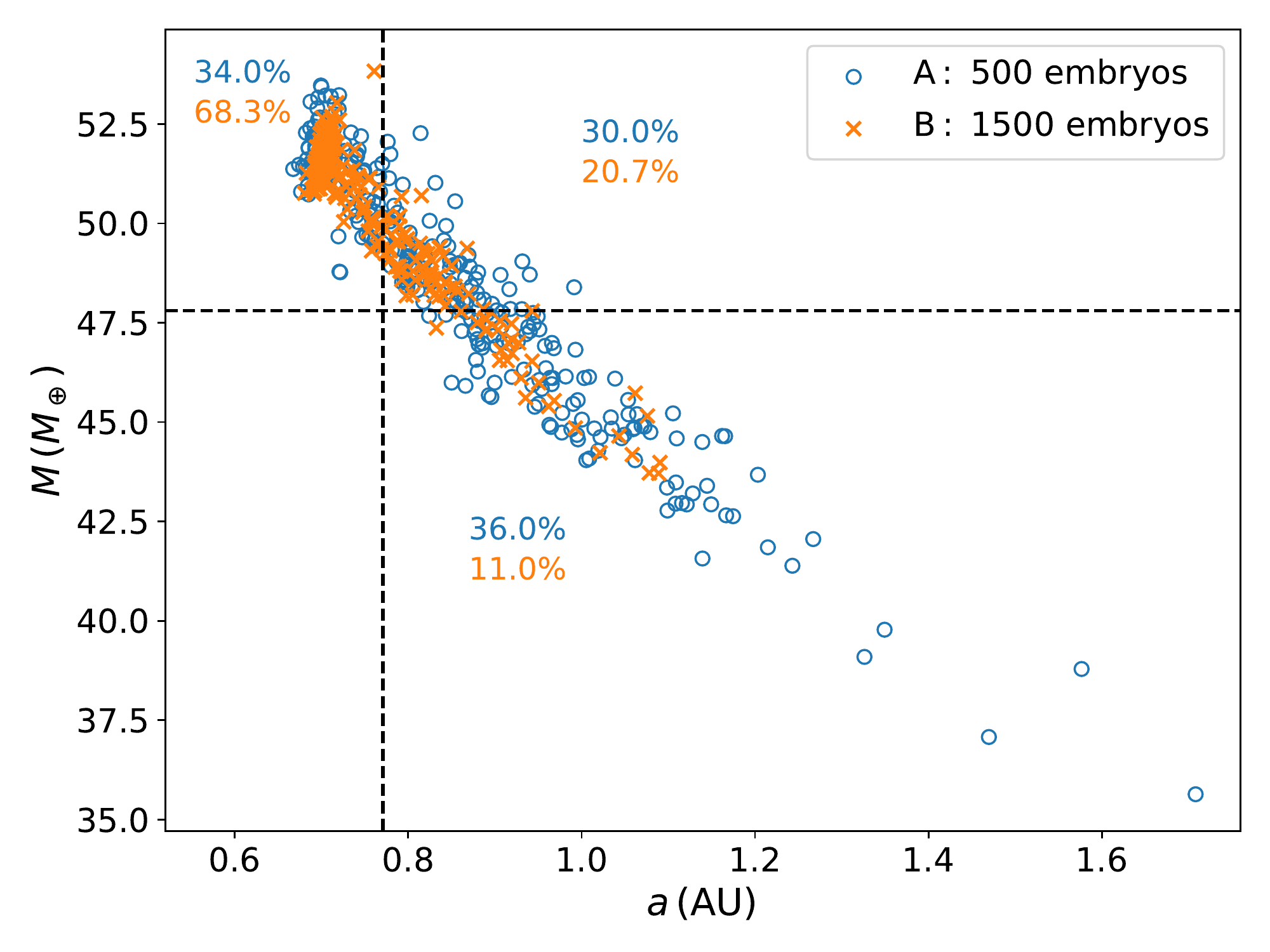}
    		\caption{The final semi-major axis and mass of 55 Cnc f at the end of integration from all 300 runs for scenarios A (500 embryos, blue) and B (1500 embryos, orange). The vertical and horizontal lines indicate the initial semi-major axis and mass of 55 Cnc f, which are 0.7708 AU and 47.8 $M_{\oplus}$, respectively. The fractions in three quadrants divided by two dashed lines are also presented.}
    		\label{fig:amf}
	\end{figure}

As the system evolves, the dynamics becomes more complicated because some protoplanets could have grown to a large size, which exceeds the threshold of mass ratio we set for PM. In this case the integrator switches to the RLM for collisions between these protoplanets and planets 55 Cnc f or d. As a result, the mass of planets decreases when the mass loss exceeds the mass of projectile protoplanets. This situation only occurs to 55 Cnc f in our simulations due to its lower mass and more collisions with other bodies (see Table~\ref{tab:mod}). In this kind of collisions with material loss, the orbit of 55 Cnc f tends to move inward because of the energy loss. Besides, the collisions of 55 Cnc f with embryos or protoplanets from its outer orbits could also slow down its outward movement caused by the gravitational attraction of the disk and even make it turn around to move inward. According to the variations (final value minus initial value) in semi-major axis $\Delta{a_f}$ and mass $\Delta{M_f}$, Fig.~\ref{fig:amf} presents three different final states of 55 Cnc f caused by above mechanisms. Obviously, in scenario B, 55 Cnc f has a higher probability to move inward due to more collisions, and also has a lower probability to lose weight because it is more difficult for less massive embryos to reach the mass ratio threshold. 

Meanwhile, 55 Cnc d could always gain materials from collisions with embryos and protoplanets as a result of PM. The final mass is around 992.4 $M_{\oplus}$ with a growth of $\sim0.75$ $M_{\oplus}$. Furthermore, the semi-major axis decreases to $\sim 5.57$ AU due to the gravitational interaction with the disk between 55 Cnc f and d. Therefore, present orbits of 55 Cnc f and d should be different from that before giant impacts. 55 Cnc d is supposed to be further away from the star 55 Cnc A before the late-stage accretion phase while the location of 55 Cnc f at that time cannot be determined by our simulations.

\begin{figure*}
	\centering
     \includegraphics[width=\hsize]{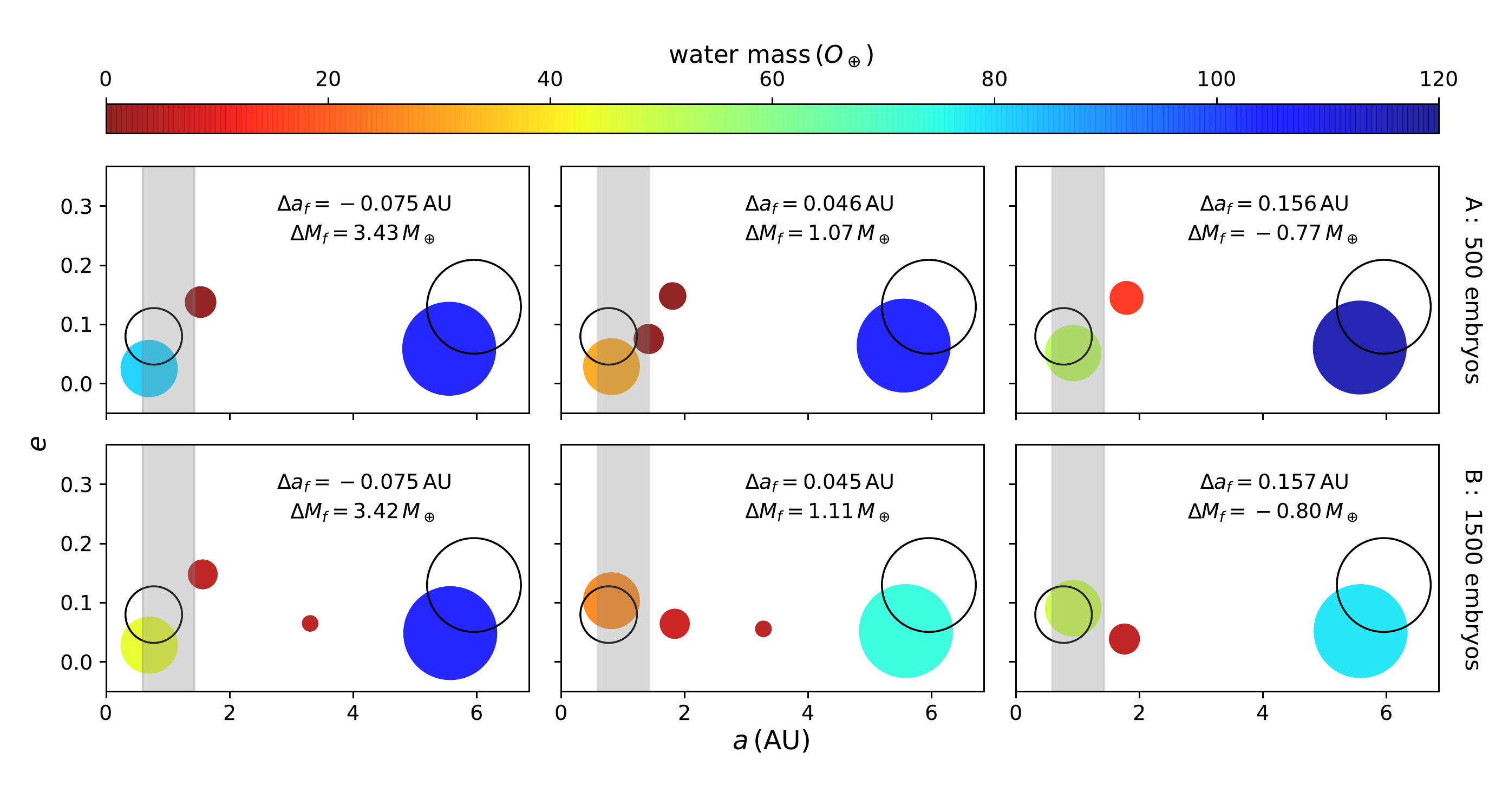}
    		\caption{Examples of final systems in scenarios A (500 embryos, \textit{upper}) and B (1500 embryos, \textit{lower}). The formed planets and 55 Cnc f,d are all shown on the $(a,e)$ plane with the size proportional to the cube root of mass and the color indicating water mass (in $O_{\oplus}$). Initial 55 Cnc f and d are represented by open circles. The grey shadow indicates the potentially habitable zone (0.59--1.43 AU). \textit{left column}: $\Delta{a_f}<0,\ \Delta{M_f}>0$, \textit{middle column}: $\Delta{a_f}>0,\ \Delta{M_f}>0$, \textit{right column}: $\Delta{a_f}>0,\ \Delta{M_f}<0$. The variations in semi-major axis and mass for 55 Cnc f are presented in each panel.}
    		\label{fig:sysexp}
	\end{figure*}

We show in Fig.~\ref{fig:sysexp} three arbitrary examples corresponding to three different final states of 55 Cnc f in scenarios A and B, respectively. Compared to the newly formed planets, 55 Cnc f and d could absorb more water, especially for the latter which is easier to accrete water-rich embryos and protoplanets. The eccentricities of planets could either be damped via dynamical friction provided by embryos or excited by energetic collisions. For 55 Cnc d, the former mechanism always dominates and causes a lower eccentricity, as we can see in Fig.~\ref{fig:sysexp}. There are always planets formed in the region between 1.5--1.9 AU. The locations of these planets are found to be somewhat related to the final semi-major axis of 55 Cnc f, which determines the stable region nearby. We note that the selected examples in each column have similar $\Delta{a_f}$ and $\Delta{M_f}$ for scenarios A and B and the final locations of formed planets are more or less similar too. We suppose that although the initial amount of embryos are different and the outcome of collisions are randomly calculated in our simulations, the planets can always form in the stable region determined by the planetary configuration, but with different mass and water content.

\subsection{Characteristics of potentially habitable planets}\label{subsec:phplanet}

\begin{figure}
	\centering
     \includegraphics[width=\hsize]{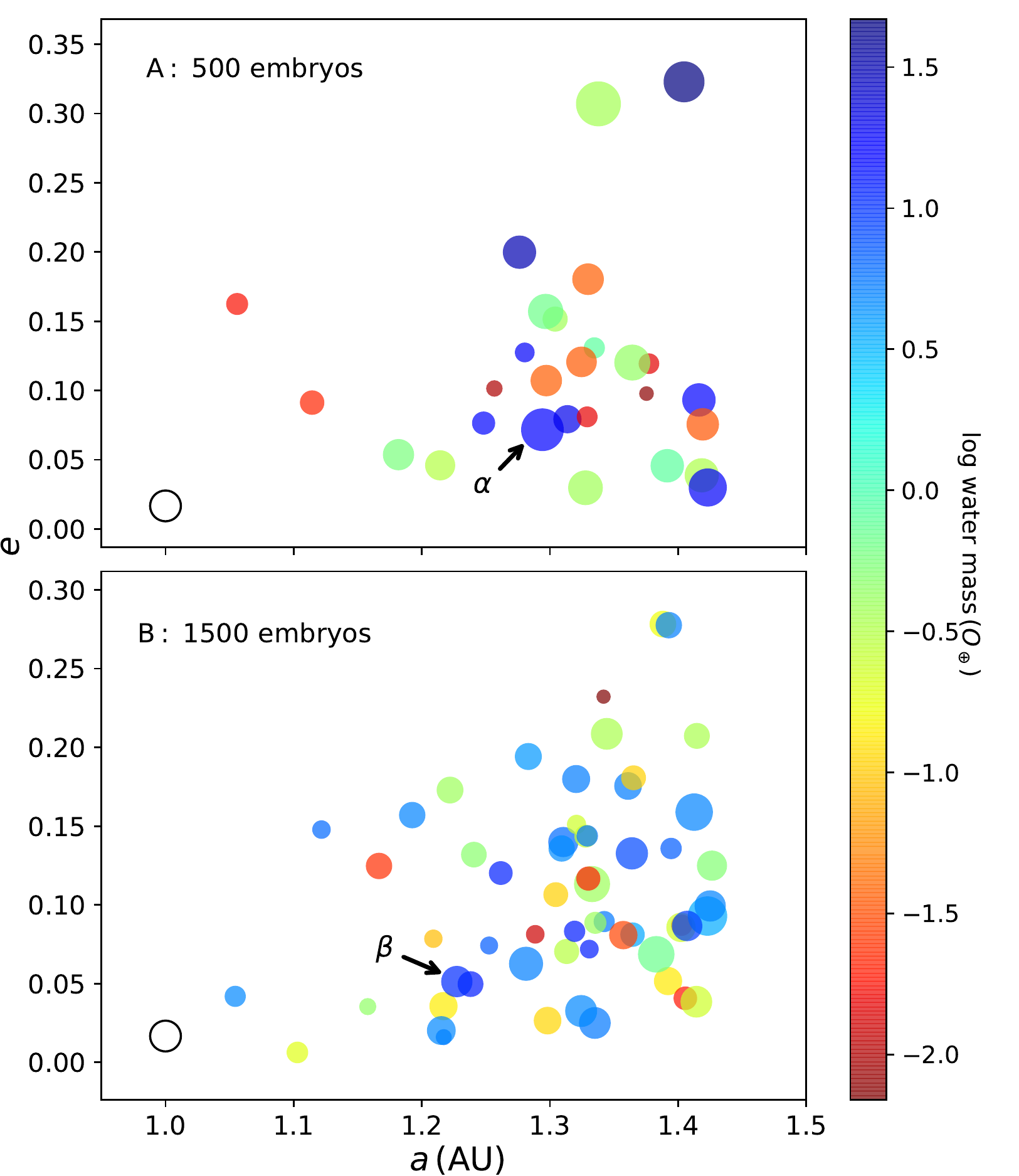}
    		\caption{Distribution of formed planets in the potentially habitable zone (0.59--1.43 AU) on the $(a,e)$ plane for scenarios A (500 embryos, \textit{upper}) and B (1500 embryos, \textit{lower}) from all 300 runs. The size is proportional to the mass while the color indicates the logarithm of water mass (in $O_{\oplus}$). The Earth is represented by an open circle. Black arrows point to planets $\alpha$ and $\beta$ of which the orbital evolution are shown in Fig.~\ref{fig:exporb}. We note that three potentially habitable planets with semi-major axis around 0.6 AU in scenario A are ignored for better visibility.}
    		\label{fig:php}
	\end{figure}
	
As we mentioned before, only 11\% and 17\% planets form in the potentially habitable zone, which account for 9\% and 14\% of the total mass and carry 11\% and 15\% of the total water inventory in formed planets in scenarios A and B, respectively (see Table~\ref{tab:mod}). The distributions of potentially habitable planets on the $(a,e)$ plane are shown in Fig.~\ref{fig:php}, from which we can see that most planets have eccentricities lower than 0.25 and many planets are clustered between $e=0.07$ and 0.14, especially for scenario B. The amount of potentially habitable planets increases with semi-major axis (also see Fig.~\ref{fig:dista}) and most of them reside beyond 1.2 AU. The inclinations are excited to 2$^\circ$--7$^\circ$ for a large portion of planets and are not higher than 17.5$^\circ$ for all potentially habitable planets in both scenarios.

In scenario A, 15 of the 31 (48\%) potentially habitable planets from 300 runs could grow to be larger than the Earth, of which the largest mass is 2.13 $M_{\oplus}$. Despite more planets in the potentially habitable zone in scenario B (58), only 12 (21\%) of them are more massive than the Earth and the largest mass is 1.63 $M_{\oplus}$. As we mentioned before, more embryos from the outer disk (beyond the snow line) have the chance to bring water to the potentially habitable zone in scenario B and thus more planets (50\% v.s. 26\%) could reserve water content more than 1 $O_{\oplus}$ finally. However, the most water-rich planets with water mass over $15\ O_{\oplus}$ (up to $\sim46\ O_{\oplus}$) are all produced by scenario A because of a higher water content of individual embryos and fewer collision loss.

\begin{figure*}
	\centering
    \includegraphics[width=0.49\hsize]{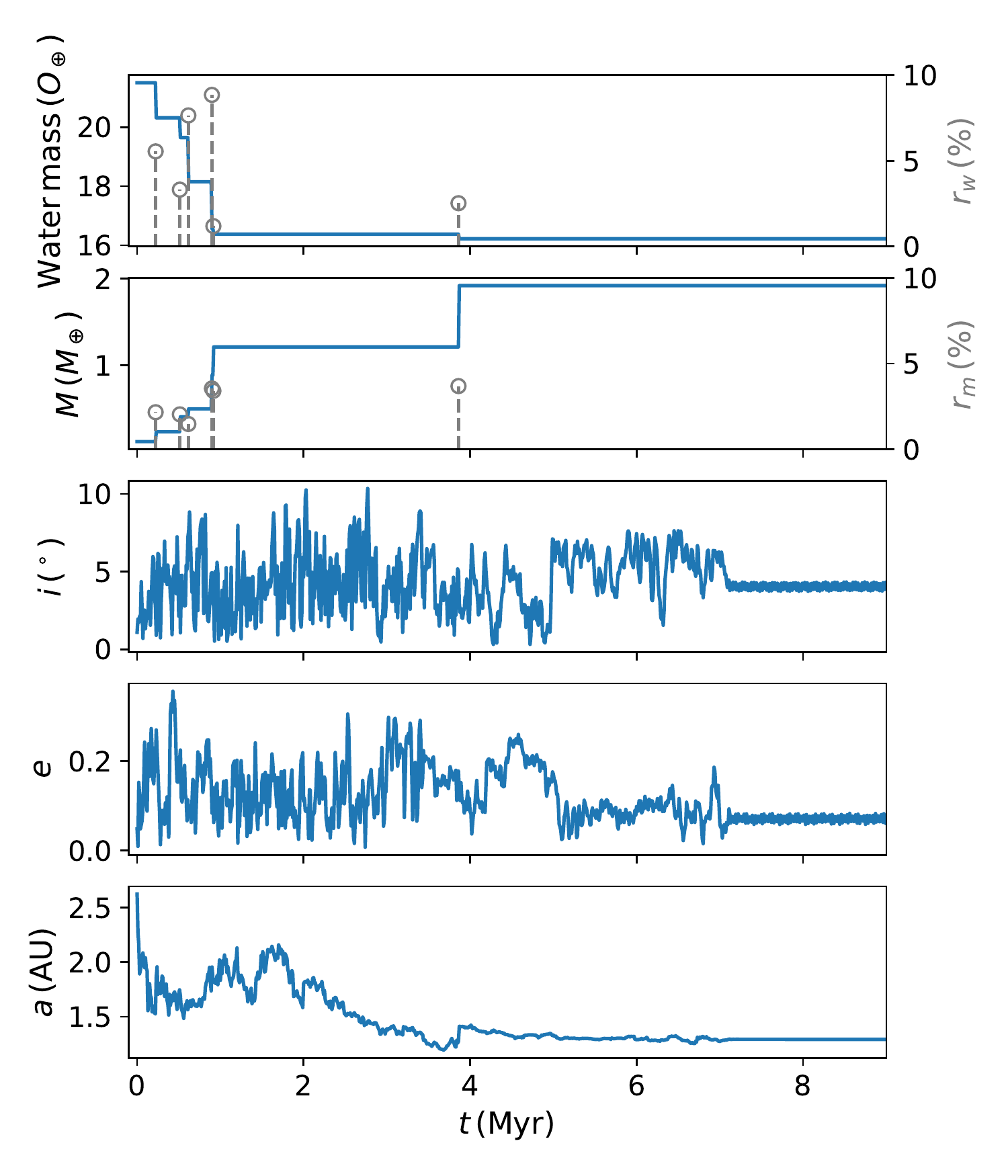}
    \includegraphics[width=0.49\hsize]{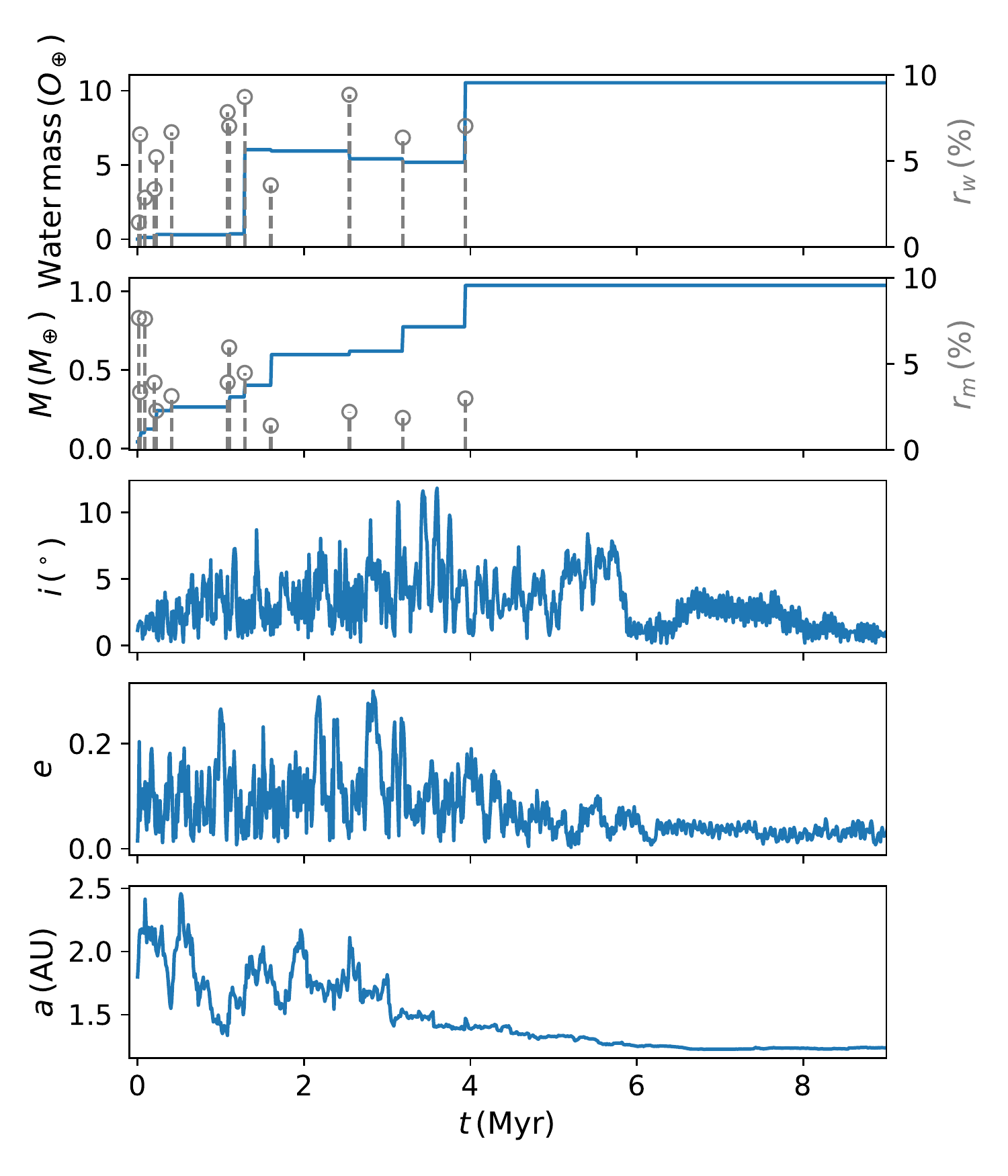}
    		\caption{Orbital evolution of two potentially habitable planets shown in Fig.~\ref{fig:php} for scenarios A (planet $\alpha$, \textit{left}) and B (planet $\beta$, \textit{right}). The grey bars indicate the loss ratio of water ($r_w$) and mass ($r_m$) in every collision.}
    		\label{fig:exporb}
	\end{figure*}

We choose two potentially habitable planets with high water inventory ($\alpha$ and $\beta$, marked in Fig.~\ref{fig:php}) from scenarios A and B, and show the orbital evolution of them in Fig.~\ref{fig:exporb}. Planet $\alpha$ originates from the outer disk with a high water mass fraction of 0.005 while planet $\beta$ originates from a water-poor region within 2 AU. In 500 embryos simulations, planet $\alpha$ was scattered inward at the beginning. It experienced six collisions in total and four of them happened within the first million year. It grew by accreting other water-poor embryos and protoplanets but also lost part of water at the same time. From 0.89 Myr, planet $\alpha$ merged with two $\sim0.4\,M_{\oplus}$ protoplanets and grew to 1.3 $M_{\oplus}$ within 30 kyr. Although it was still in a dynamically hot orbit at that time, no collision happened to planet $\alpha$ until $t=4\,{\rm Myr}$ due to the low amount of bodies then. A protoplanet of 0.8 $M_{\oplus}$ was accreted finally to make planet $\alpha$ grow to 2 $M_{\oplus}$ with water mass of 16 $O_{\oplus}$. About 3 Myr later, planet $\alpha$ entered the stable region and the eccentricity and inclination were damped to $\sim0.07$ and $\sim4^\circ$, respectively.

Planet $\beta$ is a 1 $M_{\oplus}$ planet with total water mass of 11 $O_{\oplus}$. During the formation process, planet $\beta$ encountered 13 collisions within 4 Myr, much more than planet $\alpha$ due to the larger amount of embryos. We note that at 1.28 Myr planet $\beta$ accreted a water-rich protoplanet which has moved inward to 1.55 AU and gained 5.5 $O_{\oplus}$ water from it. Another larger protoplanet also containing 5.5 $O_{\oplus}$ water was again accreted by planet $\beta$ at 4 Myr. There is another planet formed in this planetary system besides planet $\beta$. It has a slighter larger mass (1.44 $M_{\oplus}$) and resides around 2.28 AU in an eccentric orbit ($e\approx0.13$), outside the potentially zone. This planet accreted only water-poor embryos from the inner disk and is almost dry. Due to the perturbation from this additional planet, it takes much longer time ($\sim20$ Myr) for planet $\beta$ to move to stable orbits with $e=0.05$ and $i=4.9^\circ$.

\subsection{Accretion and water transport}\label{subsec:accre}

\begin{figure}
	\centering
    \includegraphics[width=\hsize]{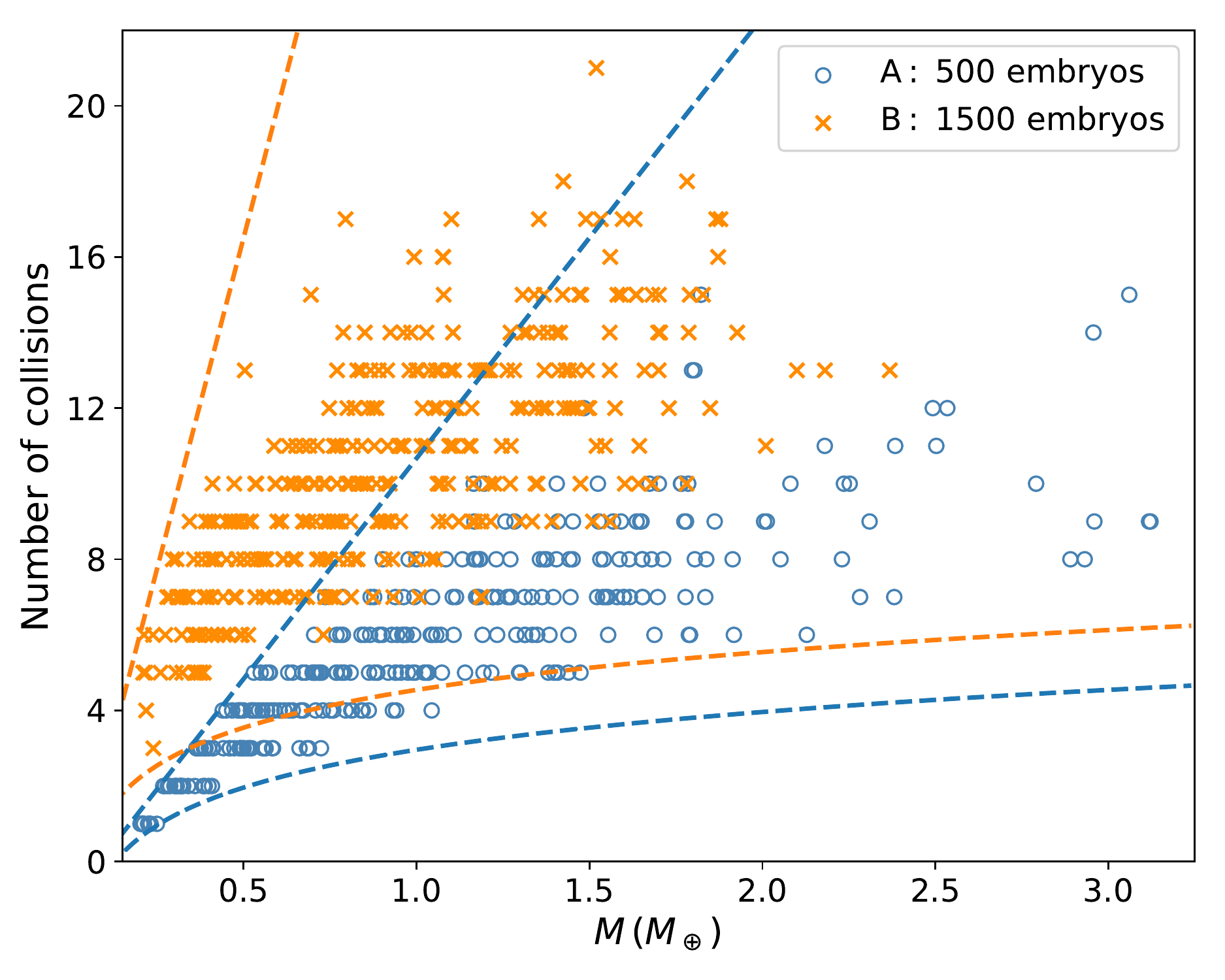}
    		\caption{The number of collisions against mass of formed planets for scenarios A (blue) and B (orange). The dashed lines depict the theoretical limit for PM.}
    		\label{fig:ncol}
	\end{figure}

\begin{figure}
	\centering
    \includegraphics[width=\hsize]{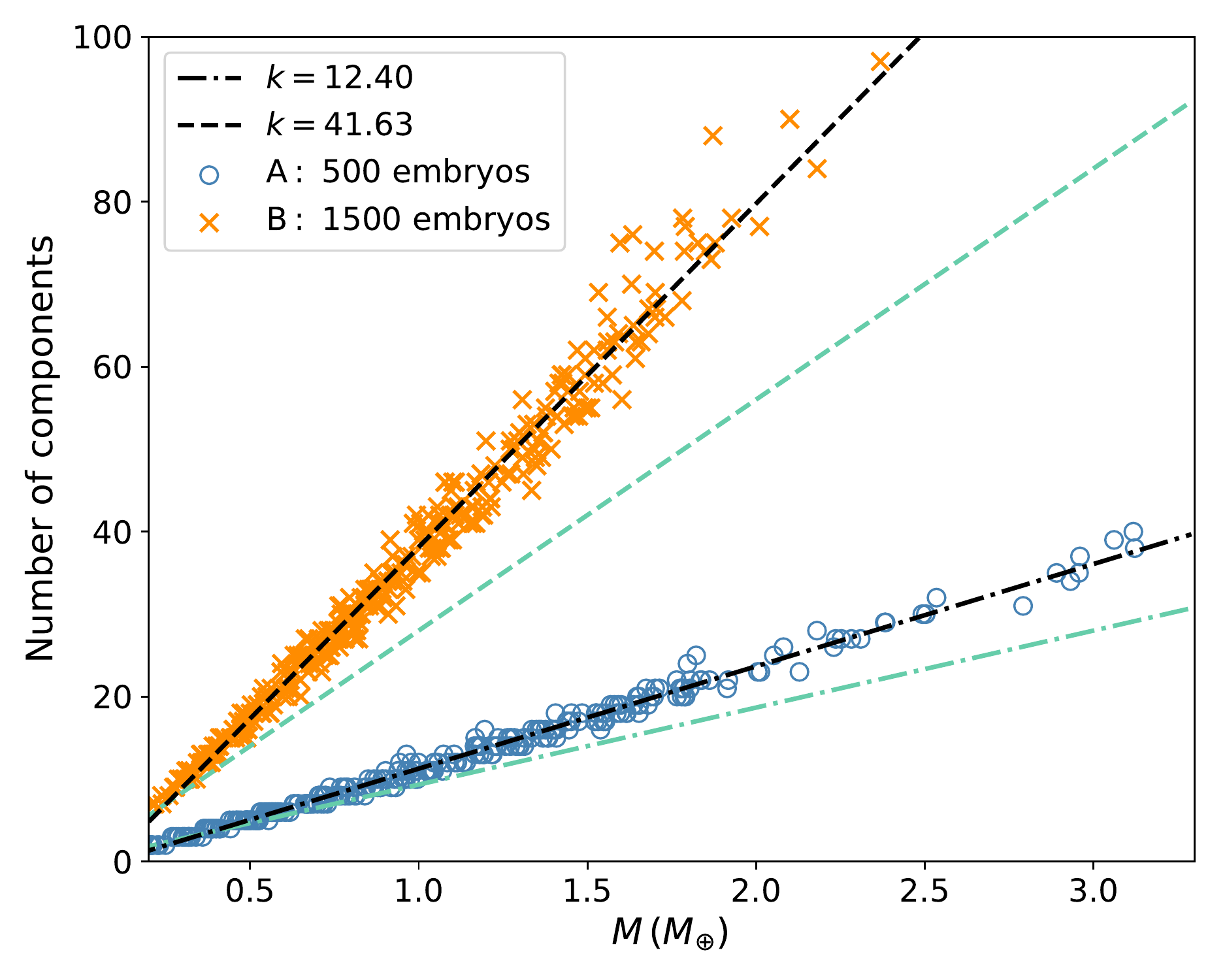}
    		\caption{Number of components against mass for scenarios A (blue) and B (orange). The green lines depict the theoretical estimation for PM while the dark lines represent the numerical linear fit.}
    		\label{fig:npiece}
	\end{figure}

As we have seen in Fig.~\ref{fig:exporb}, the total mass accreted by a planet is mainly determined by the number of collisions and the mass of projectile in each collision. On average there are 162 collisions in each simulation of scenario A, of which 118 occur among embryos and protoplanets (see Table~\ref{tab:mod}). The decay of the amount of bodies in planetary systems is mainly due to the ejections. Over 67\% embryos and protoplanets have been ejected during the integration. Therefore the formation of terrestrial planets in 55 Cnc is not efficient because most embryos are not involved in collisions and accretion. The reduction in number of embryos could reduce this proportion and only 60\% bodies have been ejected in scenario B.

In our simulations, we always store and update the data of the outcome to the most massive body in every collision. This means that during the formation of planets, the bodies that we track always merge with some other embryos or protoplanets that are less massive than themselves. When we follow the evolution of these formed planets, we can count the number of collisions they have encountered before the end of integration (see e.g. Fig~\ref{fig:exporb}). The results for all formed planets are shown in Fig.~\ref{fig:ncol}. In general the number of collisions in scenario B is obviously larger than that in scenario A. The maximum number of collisions are 15 and 21 for scenarios A and B, respectively. We can assume all accretion to be PM to approximate the upper and lower limits of the number of collisions. The upper limits are calculated by dividing the final mass of formed planets by the lowest mass of embryos. Due to the mass loss, the real upper limit should be larger. The embryos can grow most rapidly by merging with others that have the same mass. Therefore the lower limit should be $\log_2(M_{\rm f}/M_0)$, where $M_{\rm f}$ and $M_0$ are the mass of formed planets and embryos, respectively.  

To avoid considering the mass of projectiles in every single collision, we can directly count how many components of embryos each formed planet has. This so-called number of components (shown in Fig.~\ref{fig:npiece}) is always larger than the number of collisions because an embryo or protoplanet could merge with other bodies which have already accreted mass before (cf. Fig.~\ref{fig:ncol}). The number of components can be roughly estimated by dividing the final mass by initial mass for PM (green lines in Fig.~\ref{fig:npiece}). However, when considering the mass loss, more components are needed for the same mass of formed planets. We fit the number of components with a linear function and obtain the slope $k$ for scenarios A (12.40) and B (41.63), respectively. The slope representing the inverse of the mean mass of each component, could reflect the effects of the mass loss. Since $41.63/12.40\approx3.4>3$ (3 is the ratio of the embryo mass in scenarios A and B), the mass loss has a larger effect on 1500 embryos simulations (scenario B). This is obviously because more mergers result in more mass loss for all components since each component can be regarded as losing some mass in every collision. This leads to a smaller average component mass at the end and thus a lager slope.

\begin{figure*}
	\centering
    \includegraphics[width=\hsize]{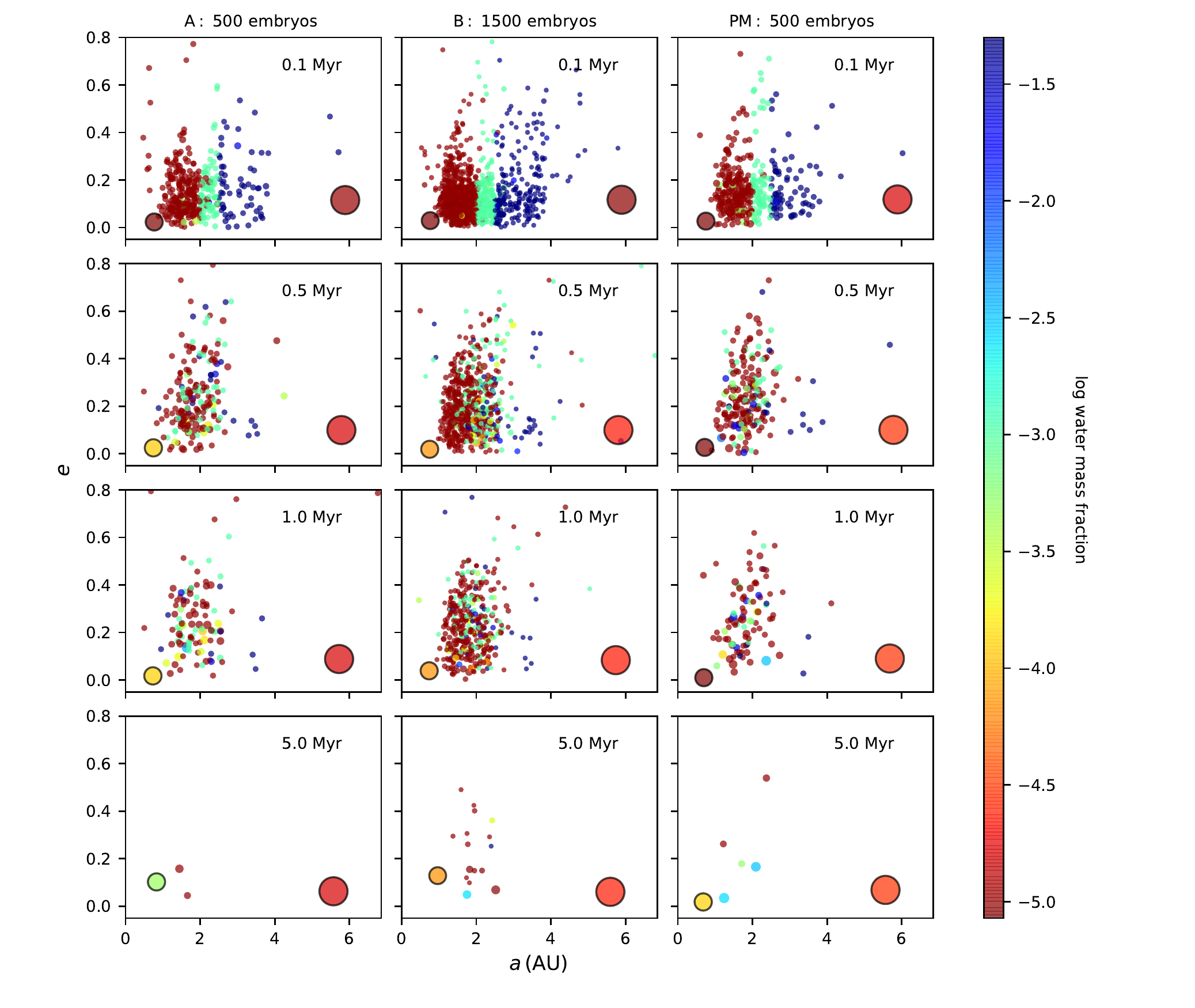}
    		\caption{Snapshots on the $(a,e)$ plane of the examples in scenarios A (\textit{left}), B (\textit{middle}) and PM (\textit{right}). The size of the symbol is proportional to the cube root of mass while the color indicates the logarithm of water mass fraction. From \textit{top} to \textit{bottom}, fours rows correspond to integration time of 0.1 Myr, 0.5 Myr, 1 Myr and 5 Myr, respectively. The planets 55 Cnc f and d are represented by circles with black edge.}
    		\label{fig:snaps}
	\end{figure*}

Given the initial water distribution and water loss parameters, the evolution of water is completely determined by the dynamical process but does not affect the dynamics in turn because the water loss is always included in the mass loss when dealing with collisions in our simulations. As we know, water-rich embryos are located beyond 2.5 AU initially while the most likely region for planet formation is between 1.5 and 2.1 AU. Hence the radial mixing plays a vital role in delivering the water from the outer disk to the inner disk. Whether there is at least one water-rich body that could cross the snow line and participate in the formation of terrestrial planets is decisive for the final water inventory of formed planets.

We show in Fig.~\ref{fig:snaps} three arbitrary examples of simulation snapshots at different time for scenarios A, B and PM. Many embryos and protoplanets have been excited to dynamically hot orbits with eccentricity over 0.2 at 0.1 Myr for all scenarios (also see Fig.~\ref{fig:exporb}). About 20\% bodies have been removed due to either collisions or ejections at that time and about 30\% of them are from the outer disk (beyond the snow line). Meanwhile, a few embryos or protoplanets from 2--2.5 AU have moved inward to the high density region and probably have collided with the water-poor bodies there. In the next 0.4 Myr, the radial mixing continues with lots of collisions taking place and the amount of bodies decays to $\sim 40\%$. The accretion of mass and water for 55 Cnc d is mainly carried out within the first half million year. During this period, many water-rich protoplanets have formed and most of them reside within 2.5 AU. However, most of these protoplanets would collide with 55 Cnc f or be ejected from the system in the next few million years. By the end of 5 Myr, only a few bodies are left in relatively stable orbits, for which it is much more difficult to accrete mass and water since then due to the low number density. Normally the last collision always happens before $\sim20$ Myr and $\sim40$ Myr for 500 and 1500 embryos simulations, respectively, but in several cases, the collisions could last to $\sim90$ Myr.

For 1500 embryos simulations, the radial mixing is more efficient and more water-rich bodies could enter the inner disk. This results in more possible water-rich planets but with lower mass and less water than those in 500 embryos simulations. The protoplanets could accrete completely via PM, which could accelerate the oligarchic growth. As we can see, the planets formed via PM are typically larger and have more water inventory (also see Table~\ref{tab:mod}). 

\subsection{Influence of different loss ratios}\label{subsec:lossratio}

Since the loss ratios are determined by the comparison between simulation results from the RLM and SPH based on the solar system (see Section~\ref{sec:compare}), they may be inappropriate for different planetary systems. The loss parameters should be related to the overall impact velocity, impact angle, mass ratio, total mass and so on. In consideration of the location of the disk, the typical orbital velocity and size of embryos, we can estimate the values of these features which determine the outcome of collisions and then get a rough material loss range by comparing these features with that for our Solar System. Furthermore, the water evolution is found much less sensitive to the water loss parameters because the initially differentiated distribution of water makes the reserved water content determined mainly by the fact that whether or not there is any water-rich embryo involved in the formation of planets rather than the water loss in each collision. For 55 Cnc, the embryo disk is assumed to extend from 1 to 4 AU. Considering the similar star mass and a closer heliocentric distance of the disk in the solar system (0.5--4 AU), the typical velocity of Kepler motion is smaller for embryos in 55 Cnc, which could result in smaller impact velocities. However, the lack of planetesimals reduces the dynamical friction in 55 Cnc and thus increases the impact velocities. On the other hand, the sizes of embryos are more comparable to each other in 55 Cnc than in the solar system and this could increase the material loss in collisions. Therefore, we can roughly estimate the collision loss parameters but it is difficult to accurately determine them.

So it is still helpful to consider different sets of loss ratios and investigate how they influence the formation of planets. We carried out 100 runs based on 500 embryos for each of three scenarios (C, D and E) adopting different sets of loss ratios. The lower limits of loss ratio are always 1\% for both mass and water. The upper limit of mass loss $\delta_m^{up}$ varies from 5\% to 15\% and are always 2\% smaller than the upper limit of water loss. Another 100 runs were conducted also for PM which corresponds to the situations where material loss is zero. The parameters and results are also summarized in Table~\ref{tab:mod}.

\begin{figure*}
	\centering
    \includegraphics[width=0.9\hsize]{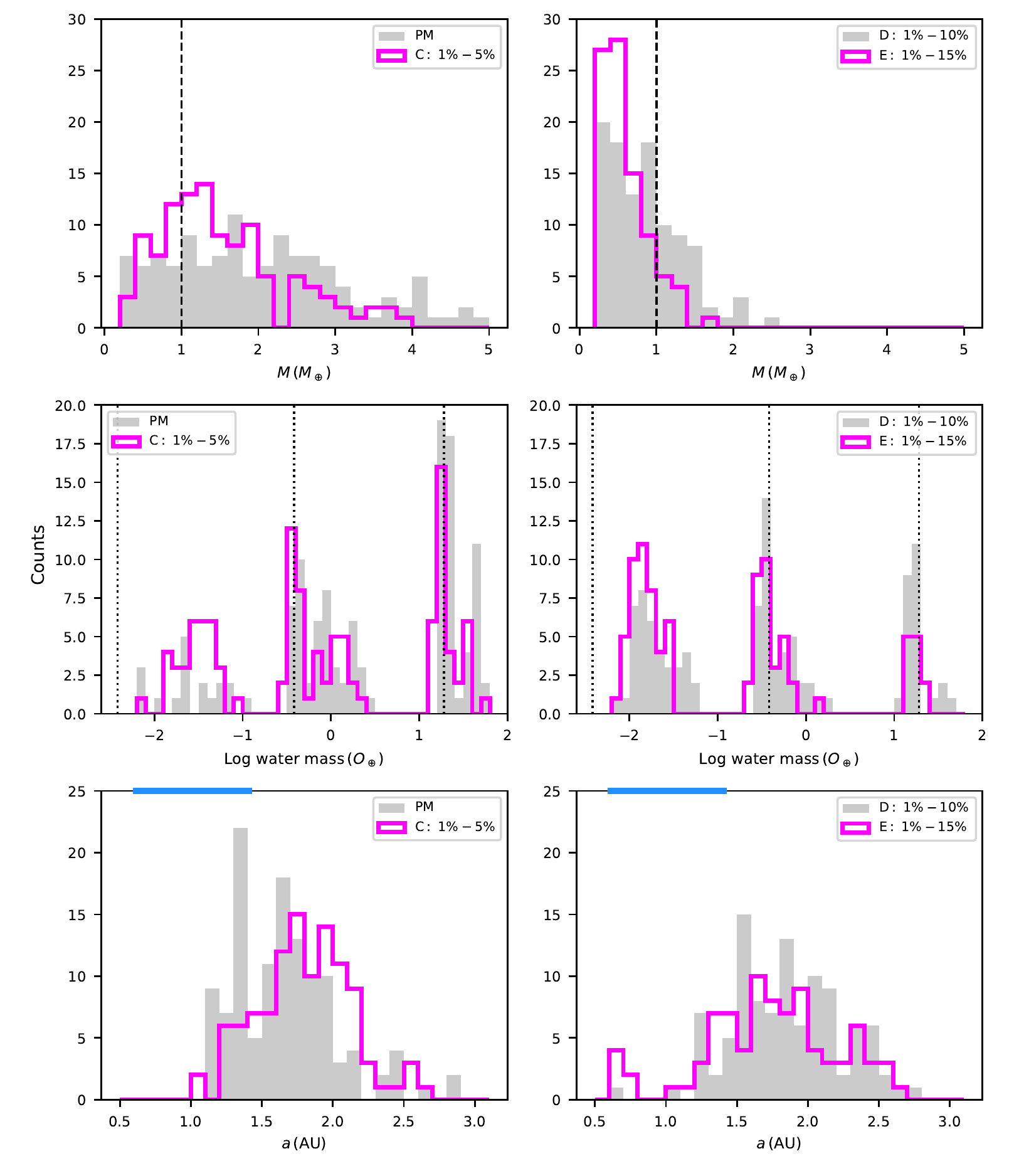}
    		\caption{Distributions of mass (\text{top}), water mass (in log, \textit{middle}) and semi-major axis (\textit{bottom}) of formed planets in scenario C, D, E and PM from all 100 runs for each scenario. The dashed lines in the \textit{top} row indicate $M=1\,M_{\oplus}$; the dotted lines in the \textit{middle} row indicate different initial water mass for 1 $M_{\mars}$ bodies with different semi-major axis; the blue bars in the \textit{bottom} row indicate the potentially habitable zone (0.59--1.43 AU).}
    		\label{fig:diffrloss}
	\end{figure*}

As we can see from Table~\ref{tab:mod}, less material loss could result in more massive planets with higher water content. Compared to our primary scenario A with mass loss ratio from 1\% to 8\%, a lower mass loss from 1\% to 5\% (scenario C) could retain 56\% more mass and 39\% more water. For PM, the increments are 133\% and 188\% in mass and water content, respectively. In scenarios D and E where loss ratios are higher, the mass of formed planets is reduced by 17\% and 51\% while the water content is reduced by 13\% and 63\%, respectively. Furthermore, the mass of potentially habitable planets also decreases with mass loss ratios. However, the water content in the potentially habitable zone is not necessarily dependent on the loss ratios and it is much more uncertain. We note that the number of collisions among embryos and protoplanets are more or less the same for different scenarios (A, C, D, E and PM), but lower loss ratio could cause more collisions to 55 Cnc f.

The final mass of formed planets for different scenarios shown in Fig.~\ref{fig:diffrloss} suggests that lower loss ratios lead to smoother mass distributions, which means more proportion of massive planets (cf. Fig.~\ref{fig:distm}). There are 79\% and 72\% planets larger than the Earth in PM and scenario C while this fraction decreases to 33\% and 11\% in scenarios D and E. The maximum mass of formed planets also decreases with mass loss ratios. The situation is similar for water content. The lower the loss ratio, the higher the proportion of water-rich planets (cf. Fig.~\ref{fig:distw}). We can still see three separate regimes in distributions of water mass (Fig.~\ref{fig:diffrloss}). The most likely region to hold planets is still located between $\sim1.5$ and $\sim2.1$ AU in scenarios C, D and E (cf. Fig.~\ref{fig:dista}). Another important reservoir is near the location of 4:1 MMR with 55 Cnc d and lots of planets are clustered around 1.3 AU when we adopt PM. Therefore the planets have a much higher probability (31\%) of forming in the potentially habitable zone via complete accretion (also see Table~\ref{tab:mod}).

We note that the uniform distribution of material loss we adopt could result in different results compared to just using the expectation value all the time (e.g. 1\% -- 8\%, and 4.5\%). We also considered different ranges of loss ratios with the same expectation values and the results show that the upper limit of the loss ratios has more effects on the formed planets than the lower limits.

\section{Conclusions and discussions}\label{sec:condis}

In this paper we propose the random loss method (RLM) to deal with collisions of celestial bodies during the formation of planetary systems. Unlike perfect merging (PM), part of the mass and water content would be subtracted from the sole outcome of collisions. Given appropriate loss ratios, the material loss in each collision is randomly determined in a range dependent on the total mass and water inventory of colliding bodies. Based on a sufficient number of simulations, the statistical method could provide meaningful and convincing results.

To verify the feasibility and accuracy of the RLM, we apply it to the planet formation in the solar system and compare the results with SPH simulations from \citet{2020A&A...634A..76B}. With appropriate loss parameters selected, i.e. from 1\% to 8\% for the mass loss and from 1\% to 10\% for the water loss, the main conclusions including the total mass of formed planets ($\sim$1.48 $M_\oplus$) and the mass in the potentially habitable zone ($\sim$0.93 $O_\oplus$) are consistent with each other for both methods. Although fewer planets formed via the RLM (1.79 v.s. 2.40), the number of potentially habitable planets is more or less the same (0.96 v.s. 1.00) in our solar system. Furthermore, the distribution of the formed planets on the $(a,M)$ plane with a peak around 1.35 AU for the RLM agrees well with that for SPH, which is a great improvement compared to PM. 

In spite of fewer collisions and less material loss in simulations with the RLM, the average mass and water loss per collision for the RLM and SPH are very similar ($\sim0.01\,M_\oplus$ for mass loss and $\sim0.088\,O_\oplus$ for water loss). Despite a smaller mean value of the final water content in our simulations, the difference from SPH results is still in an acceptable range due to the fact that this formation process is highly chaotic. Neither PM nor the RLM could reserve as much water as that in SPH and we suppose that hit-and-run, in which still two bodies are retained after collisions, could enhance the radial mixing and thus efficiently deliver the water in giant impacts. Besides, the frequent occurrence of hit-and-run in SPH simulations ($>50\%$) could relieve the decay of the body amount and is detrimental to form large planets. The average mass of formed planets is only 0.61 $M_\oplus$ for SPH while it is 0.83 $M_\oplus$ and 1.00 $M_\oplus$ for the RLM and PM, respectively. It is evident that PM overestimates the amount and mass of formed planets. Furthermore, the deviations of results statistically obtained via the RLM with different initial conditions are much smaller than that from the other two methods. To sum up, the RLM is a promising approach which can be simply implemented and requires much less computational resource than SPH does.

We aimed to use the RLM to study the planet formation between two massive planets. Due to the similarity to the solar system, we choose the extrasolar planetary system 55 Cnc as a case, which hosts (at least) five planets with a gap between 55 Cnc f ($a=0.7708$ AU) and d ($a=5.957$ AU). In this paper we focus on the questions: whether or not any planet could form between 55 Cnc f and d, if any, what is the probability, what are the characteristics of formed planets and how they formed? 

We assume a 500 Mars mass embryo disk between $\sim1$ and $\sim4$ AU. The initial radial distribution is numerically generated from the time duration distribution of massless particles moving in the gravitational field of the star and all five planets. For embryos we adopt the water mass fraction that is widely used in the case of the solar system, where the water-rich objects all reside beyond the snow line (2.5 AU). In addition to the star and embryos which are in dynamically cold orbits initially with the mass around 1 or 1/3 Mars mass in different scenarios, only two planets, 55 Cnc f and d, are included in our simulations of giant impacts. The masses of these two planets we use in our simulations are actually the minimum mass obtained from radial velocity measurements and thus the inclinations are both assumed to be zero. The formation probability and distribution of additional planets may be different if the real masses of 55 Cnc f and d could be determined by further observations (such as by transit). We allow all types of bodies (embryo, protoplanet, planet) to gravitationally interact with each other and consider the real (physical) collisions between them. 

Regardless of whether the initial number of embryos is 500 or 1500, statistical results from 300 simulation runs show that it is very likely that there exists an Earth-like planet with $\sim4$--6 Earth ocean around 1--2.6 AU. However, the most possible region is between 1.5--2.1 AU and the formed planet only has a $\sim10\%$ probability to be found in the potentially habitable zone (0.59--1.43 AU). Larger amount of embryos could make more water-rich bodies cross the snow line and deliver the water content to the inner disk. As long as one embryo from the outer disk (beyond 2.5 AU) participates in the formation of the final planet, the planet would become water-rich because at least $\sim46\%$ of initial water inventory in that embryo could still remain even after 15 collisions. 

Many work has been dedicated to investigate the possible existence of an additional terrestrial planet between 55 Cnc f and d \citep[see e.g.][]{2003AsBio...3..681V,2008ApJ...689..478R,2019PASJ...71...53S}. \citet{2012PASJ...64...73C} suggests that a possible planet may exist around 1.5 AU. \citet{2008ApJ...689..478R} found a large reservoir for additional planets between 0.9 and 3.8 AU and the stability analysis from \citet{2019PASJ...71...53S} suggests a stable region between 1 and 2 AU. Our work describes a concrete mechanism of planet formation between two massive planets and the results are consistent with the stability analysis from previous work.

Apart from mass loss ratios from 1\% to 8\%, we also adopt other loss parameters considering the differences between different planetary systems. Although lower loss ratios could result in more planets with higher mass, the possible formation region is more or less the same. The stable region is related to the evolution of planets 55 Cnc f and d, of which the mass, compositions, and orbits are also affected by the giant impacts and the interaction with the disk. The resulted planetary migration at this stage should be common in the formation of planetary systems. 

Therefore the choice of appropriate loss parameters is very important to get reasonable results. Of course we can decide them based on the typical impact velocity, mass ratio, escape velocity and so on for different systems, but it is still arbitrary. On the other hand, we always get only one outcome from collisions, just like in PM. As we mentioned, a large proportion of collisions are hit-and-run, which keeps the number of bodies unchanged and relieves the decay of the number of objects. Hence for the RLM, the total number of collisions is still somewhat underestimated.

\citet{2011ApJ...734L..13P} have demonstrated that the habitable zone for planets retaining primordial H$_2$ envelopes could be significantly extended due to the Hydrogen greenhouse warming compared to the planets with CO$_2$ atmosphere. In this case, the possible planets between 55 Cnc f and d have a higher probability to be habitable considering that the most likely region of additional planets (1.5--2.1 AU) is beyond the classical habitable zone (0.59--1.43 AU).

Unlike previous work such as \citet{2016ApJ...821..126Q,2020A&A...634A..76B}, we do not include planetesimals in our simulations. The small mass planetesimals could provide effective dynamical friction to reduce the impact energy, leading to less material loss. Moreover, planetesimals could also play a role in water delivery because they could enhance the radial mixing. Besides, \citet{2020MNRAS.496.4979D} have demonstrated that comets could also deliver water to the habitable zone in 55 Cnc via collisions. We adopt the same initial water distribution widely used for the solar system, assuming a same snow line at 2.5 AU because the G8 V star 55 Cnc A has a similar mass to the Sun and thus may be in a similar state to the Sun at the beginning of the late-stage accretion phase. However, this may not be the case considering the possibly different physical parameters such as the metallicity and  optical depth of the protoplanetary disk and a different water distribution may result in different final water inventory in formed planets. Nevertheless, keeping in mind the very similar results for the formation of terrestrial planets with two different methods (RLM and SPH) in our solar system, we are convinced to model a somewhat realistic formation process of possible terrestrial planets also in 55 Cnc.

\section*{Acknowledgements}

This work has been supported by National Natural Science Foundation of China (NSFC, Grants No. 11473016, No. 11933001) and National Key R\&D Program of China (2019YFA0706601). This research is also supported by the Austrian Science Fund (FWF) through grant S11603-N16 (R. D. and L. Z.). Lei Zhou acknowledges the financial support from the program of China Scholarship Council (No. 201906190106) for his visit in University of Vienna.

\section*{Data Availability}

The data underlying this article will be shared on reasonable request to the corresponding author. The data of SPH simulation used for comparison in Section~3. are available in the published article of \citet{2020A&A...634A..76B}, at \url{https://doi.org/10.1051/0004-6361/201936366}.


\bibliographystyle{mnras}
\bibliography{PF2} 








\bsp	
\label{lastpage}
\end{document}